\documentclass{aa}

\usepackage{graphicx}
\usepackage{subcaption}
\usepackage{multirow}
\usepackage{txfonts}
\newcommand\Mstars{$M_\star$}

\begin{document}

   \title{Possible environmental quenching in an interacting little red dot pair at $z\sim7$}
   \titlerunning{Possible environmental quenching in an interacting LRD pair at $z\sim7$}
   \authorrunning{Rosa M. M\'erida et al.}
   \author{Rosa M. M\'erida,
   \inst{1}
          Gaia Gaspar,
          \inst{1,2}
          Marcin Sawicki, 
          \inst{1}
          Yoshihisa Asada, 
         \inst{1,3}
          Guillaume Desprez, \inst{4}
          Gregor Rihtar\v{s}i\v{c},\inst{5} 
          Jacqueline Antwi-Danso, \inst{6}
          Roberta Tripodi, \inst{5,7}
          Chris J. Willott, \inst{8}
          Maru\v{s}a Brada\v{c}, \inst{5,9}
          Gabriel B. Brammer, \inst{10,11}\\
          Kartheik G. Iyer, \inst{12}
          Nicholas S. Martis, \inst{5}
          Adam Muzzin, \inst{13}
          Ga\"el Noirot, \inst{14}
          Ghassan T. E. Sarrouh, \inst{13}
          Vladan Markov \inst{5}}

   \institute{
                Institute for Computational Astrophysics and Department of Astronomy and Physics, Saint Mary's University, 923 Robie Street, Halifax, NS B3H 3C3, Canada; 
   \email{Rosa.MeridaGonzalez@smu.ca}
   \and  Observatorio Astronómico de Córdoba, Universidad Nacional de Córdoba, Laprida 854, X5000, Córdoba, Argentina 
    \and Department of Astronomy, Kyoto University, Sakyo-ku, Kyoto 606-8502, Japan 
    \and Kapteyn Astronomical Institute, University of Groningen, P.O. Box 800, 9700AV Groningen, The Netherlands
    \and Faculty of Mathematics and Physics, Jadranska ulica 19, SI-1000 Ljubljana, Slovenia
    \and David A. Dunlap Department of Astronomy and Astrophysics, University of Toronto, 50 St. George Street, Toronto, Ontario, M5S~3H4, Canada
    \and INAF - Osservatorio Astronomico di Roma, Via Frascati 33, Monte Porzio Catone, 00078, Italy
    \and National Research Council of Canada, Herzberg Astronomy \& Astrophysics Research Centre, 5071 West Saanich Road, Victoria, BC, V9E 2E7, Canada
    \and Department of Physics and Astronomy, University of California Davis, 1 Shields Avenue, Davis, CA 95616, USA
    \and Cosmic Dawn Center (DAWN), Denmark
    \and Niels Bohr Institute, University of Copenhagen, Jagtvej 128, DK-2200 Copenhagen N, Denmark
    \and Columbia Astrophysics Laboratory, Columbia University, 550 West 120th Street, New York, NY 10027, USA
    \and Department of Physics and Astronomy, York University, 4700 Keele St. Toronto, Ontario, M3J 1P3, Canada
    \and Space Telescope Science Institute, 3700 San Martin Drive, Baltimore, Maryland 21218, USA\\
    }

   \date{Received September 15, 1996; accepted March 16, 1997}

% \abstract{}{}{}{}{} 
% 5 {} token are mandatory
 
  \abstract 
  % context heading (optional)
  % {} leave it empty if necessary  
   {We report the discovery of a $z\sim7$ group of galaxies that contains two little red dots (LRDs) just 3.3 kpc apart, along with three potential satellite galaxies, as part of the Canadian NIRISS Unbiased Cluster Survey (CANUCS). The spectral energy distributions (SEDs) of this LRD pair show evidence of a Balmer break, consistent with a recent ($\sim 100$~Myr) quenching of star formation. In contrast, the satellites are compatible with a recent-onset ($\sim 100$~Myr), ongoing burst of star formation. LRD1's SED is consistent with a dust-free active galactic nucleus (AGN) being the source of the UV excess in the galaxy. The optical continuum would be powered by the emission from an obscured post-starburst and the AGN at a subdominant level. LRD2's SED is more ambiguous, but it could also be indicative of a dust-free AGN. In this scenario, these LRDs would be massive ($M_\star\sim10^{10}\,M_\odot$) and dusty (A(V) $>1$ mag) and the three satellites would be lower-mass objects ($M_\star\sim10^{8-9}\,M_\odot$) subject to low dust attenuations. The proximity of the two LRDs suggests that their interaction is responsible for their recent star formation histories, which can be interpreted as environmental bursting and quenching in the epoch of reionization.
}

   \keywords{Galaxies: interactions -- active -- high-redshift -- evolution
               }

   \maketitle
%
%-------------------------------------------------------------------

\section{Introduction}
\label{sec:intro}

Little red dots (LRDs; \citealt{Labbe2023}, \citealt{Barro2024}, \citealt{Greene2024}, \citealt{Matthee2023}) represent one of the most remarkable discoveries made by the JWST \citep{Gardner2023}. These red, compact sources show a characteristic V-shaped spectral energy distribution (SED) consisting of a nearly flat to blue rest-frame ultraviolet (UV) continuum and a very steep slope in the optical. These properties make LRDs a challenge for SED modeling, even with the aid of spectroscopic data.

As summarized in \citet{Perez-Gonzalez2024}, there are different potential explanations for LRDs, including high equivalent width ($1,000\,\AA$) emission lines that boost the broadband photometry (e.g., \citealt{Yuma2010}, \citealt{Perez-Gonzalez2023}, \citealt{Desprez2024}, \citealt{Hainline2024arXiv}); a flexible treatment of the dust attenuation (e.g., \citealt{Akins2023}, \citealt{Barro2024}, \citealt{Zhengrong2024}); active galactic nuclei (AGNs; e.g., \citealt{Inayoshi2024}, \citealt{Zhengrong2024}); and hybrid models that combine AGNs and stars (e.g., \citealt{Kocevski2023}, \citealt{Greene2024}, \citealt{Kokorev2024}, \citealt{Tripodi2024}, \citealt{Wang2024}). However, none of these approaches has been able to fully reproduce the SEDs of the bulk of the LRD population.

Little red dots  are generally X-ray-weak (\citealt{Ananna2024}, \citealt{Maiolino2024}), mostly radio-quiet (\citealt{Labbe2023}, \citealt{Perger2024}), and remain undetected in the \mbox{(sub)}millimeter range (\citealt{Casey2024}, \citealt{Labbe2025}). They do not show signs of variability \citep{Zhang2024}, and their mid-infrared SED is more consistent with emission from stars (\citealt{Perez-Gonzalez2024}, \citealt{Williams2024}). On the other hand, $>50$\% of the photometrically selected LRDs display broad Balmer and/or Paschen lines (e.g., \citealt{Kocevski2023}, \citealt{Greene2024}, \citealt{Tripodi2024}). The prevailing view is that the SEDs of these galaxies are not solely driven by an AGN, but rather have a hybrid origin. 

The formation and evolution of these galaxies is also uncertain. They have been proposed to both be the progenitors of the first massive galaxies (e.g., \citealt{Setton2024}, \citealt{Tripodi2024}, \citealt{Wang2024}) and low-mass objects hosting an AGN \citep{Chen2024}. Most LRDs are found isolated in the field, yet JWST studies reveal a mean number of merger events per galaxy of $\sim4$ Gyr$^{-1}$ at $6.5<z<7.5$ \citep{Duan2024}.  \citet{Tanaka2024} reported three of the first dual LRD candidates, with projected separations of $0.2-0.4$\arcsec, in the COSMOS-Web survey \citep{Casey2023}. Interactions could thus play a major role in the triggering of potential AGN activity and quenching in LRDs (see also \citealt{Greene2024}, \citealt{Labbe2024}).

We explore the impact of the environment on LRDs, presenting a dual LRD candidate at $z\sim7$ that could be undergoing merging, bursting, and quenching events. The pair is likely embedded in a galaxy group that has three satellite galaxies and shows signs of coordinated star formation histories (SFHs).

Throughout this work we assume $\Omega_\mathrm{M,0}=0.3$, $\Omega_{\Lambda,0}=0.7$, and H$_0=70$ km s$^{-1}$ Mpc$^{-1}$ and use AB magnitudes \citep{Oke1983}. All stellar mass ($M_\star$) and star formation rate (SFR) estimates assume a \citet{Chabrier2003} initial mass function (IMF).

\section{Data}
\label{sec:data}

We used data from the Canadian NIRISS Unbiased Cluster Survey (CANUCS; GTO program
\#1208; \citealt{Willott2022}), which consists of Near Infrared Camera (NIRCam; \citealt{Rieke2023}) and JWST Near InfraRed Imager and Slitless Spectrograph (NIRISS; \citealt{Doyon2023}) observations of five strong lensing clusters and flanking fields. CANUCS also incorporates Near Infrared Spectrograph (NIRSpec; \citealt{Jakobsen2022}) PRISM spectroscopy in these fields.  

Our objects are located behind the $z=0.375$ cluster Abell 370 and were observed with the NIRCam $F090W$, $F115W$, $F150W$, $F200W$, $F277W$, $F356W$, $F410M$, and $F444W$ filters for 6.4 ks each, reaching 3$\sigma$ point source limiting magnitudes
ranging from 29.5 to 30.2 mag. Archival \textit{Hubble} Space Telescope (HST) imaging from the Advanced Camera for Surveys (ACS; namely, $F606W$ and $F814W$) in this field were also included (HST-GO-15117 PI Steinhardt; \citealt{Steinhardt2020}). The 3$\sigma$ point source limiting magnitudes, based on the uncertainty of nearby detections, reach $\sim 27.4 - 28.0$ mag for these ACS images. Spectroscopy from NIRISS or NIRSpec is not available for these objects.

\begin{figure*}[htp]
    \centering
    \begin{minipage}{0.7\textwidth}
        \centering
        \includegraphics[width=\linewidth]{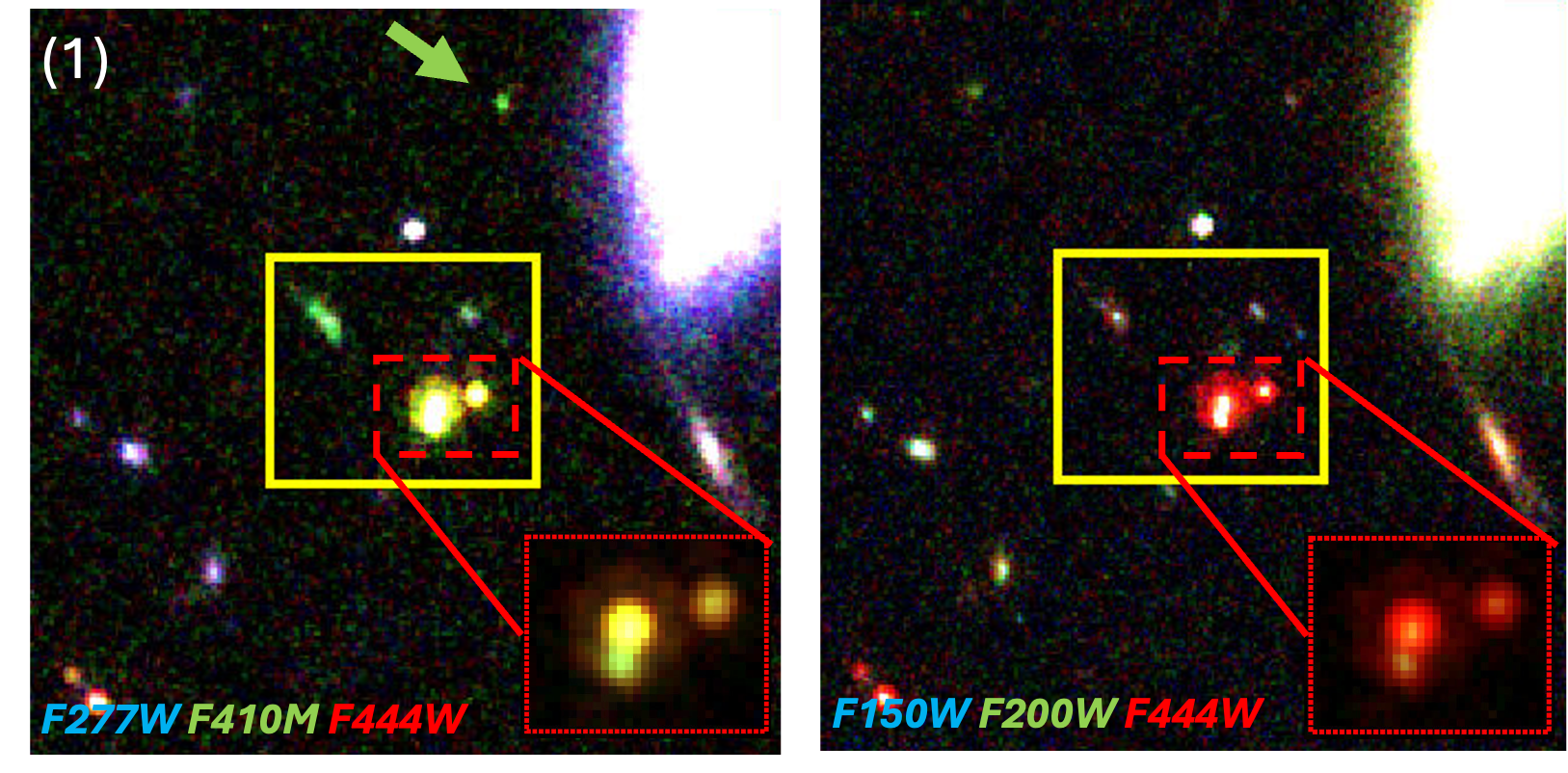}
    \end{minipage}%
    
    \begin{minipage}{0.9\textwidth}
        \centering
        {\includegraphics[width=\linewidth]{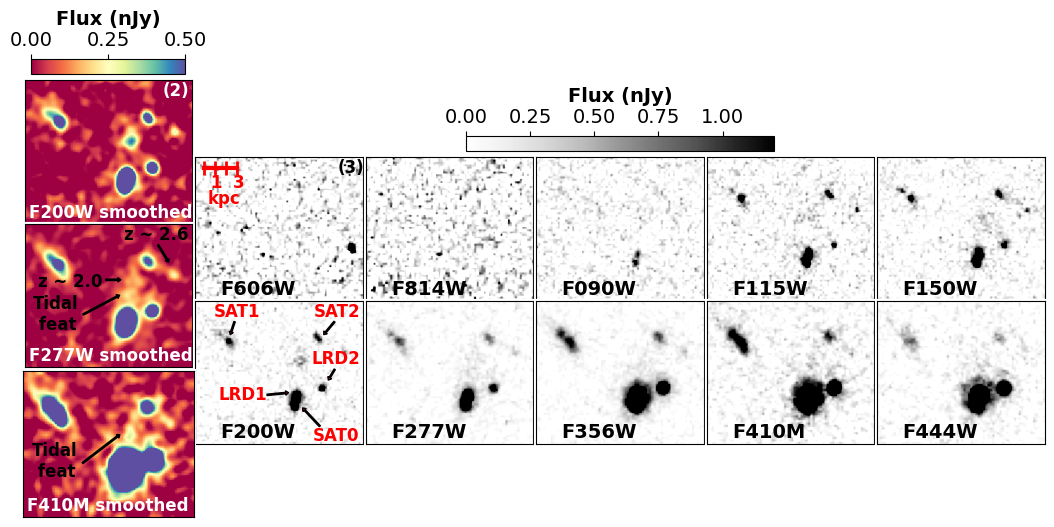}}
    \end{minipage}
    \caption{Cutouts of the galaxy group in different bands.  (1): 8$\times8$ arcsec$^2$ cutouts of two RGB images based on $F277W$, $F410M$, and $F444W$ (left) and $F150W$, $F200W$, and $F444W$ (right) NIRCam imaging. The yellow square encloses the region that contains our galaxies; its dimensions are 2.8$\times2.4$ arcsec$^2$. These images highlight the common $F410M$ excess (and thus the high-$z$ nature) in this galaxy group compared to sources in the neighborhood (left) and the more prominent emission in $F444W$ of the LRD pair compared to the satellites (right). The red squares in both panels show a zoomed-in view of the LRD pair in a less saturated scale to highlight the difference in color between the LRDs and SAT0. The arrow in the left panel points to a $z\sim7$ source $\sim18$ kpc away from the LRD pair, with colors similar to the satellite galaxies. (2): Cutouts of our galaxy group in smoothed versions of $F200W$, $F277W$, and $F410M$ obtained using a Gaussian filter with $\sigma=1.2$ pixels. (3): Postage stamps of these sources in the different HST and NIRCam bands at native spatial resolution. NIRCam images are plotted following the same scale as above. For the HST bands, the upper limit was set to 3 nJy to deal with the noise. The kiloparsec scale is not corrected for lensing magnification, which is also low at this position and redshift ($\mu = 1.26$; see Fig. \ref{fig:lensing}).
    }
    \label{fig:images}
\end{figure*}

The CANUCS and archival images were processed and bright cluster galaxies were removed as described in \citet{Noirot2023} and \citet{Martis2024}. Their point spread functions (PSFs) were then degraded to match that of the $F444W$ image \citep{Sarrouh2024}, after which photometry was done in matched apertures magnitudes \citep{Asada2024pairs}; for some of the galaxies targeted in this paper, we also performed point-source photometry (see Sect.~\ref{sec:PSFphotometry}). Photometric redshifts for all detected sources were measured with \texttt{EAzY} \citep{Brammer2008} using the standard templates augmented with the \citet{Larson2023} set (see \citealt{Asada2024photoz} for more details). We used
the intergalactic medium (IGM) attenuation curve of \citet{Asada2024photoz}, which has been shown to greatly reduce photo-$z$ systematics at $z>6$ by accounting for the effect of the Ly$\alpha$ damping wing from the circumgalactic medium (CGM) gas. 

\section{The $z\sim7$ group}
\label{sec:analysis}

\subsection{Discovery and basic properties}
\label{sec:basic_proprerties}

Examining the CANUCS catalogs, we find a potential close group of four $6.79 < z_{\mathrm{phot}}< 6.87$ objects ($z\sim7.0$ using the \citealt{Inaoue2014} standard IGM transmission in the case of the LRDs); see Fig.~\ref{fig:images}. 

In addition to the setting described in the previous section, we ran \texttt{EAzY} using the template model built from the J0647$\_$1045 LRD at $z = 4.53$ from \citet{Killi2024}. This galaxy was well fitted assuming that the optical/near-infrared emission and the broadband lines are generated from a relatively massive AGN while the less obscured UV continuum and narrow lines are produced, at least partly, by a small but spatially resolved star-forming host galaxy. The photo-$z$s found for the LRD pair are then 7.01 and 7.00, respectively.

With $\Delta z/(1+\bar{z})$=0.0229, the redshifts of all these galaxies are indistinguishable given our photo-$z$ uncertainties \citep{Asada2024pairs, Asada2024photoz}. The photo-\textit{z}s are well constrained by excess flux in the $F410M$ medium band (due to redshifted H$\beta$ and [OIII]$\lambda\lambda 4959, 5007$ emission lines), a red $F277W-F356W$ color (due to either the Balmer break or strong emission lines at $\lambda_{\rm rest} > 4000$\AA), and a sharp drop across $F090W-F115W$ (due to the Lyman Break at rest-1215\AA).

Visual inspection of the short wavelength bands reveals that the brightest source consists of two distinct components -- dubbed CANUCS-A370-LRD1 and CANUCS-A370-SAT0 (LRD1 and SAT0 for short; see the $F200W$ panel of Fig.~\ref{fig:images},  including the labels), blended in the CANUCS catalog. Using \texttt{SExtractor} \citep{Bertin1996} on the original (unsmoothed) $F150W$ image, we recovered the centroids of the five objects in the group: CANUCS-A370-LRD1, -SAT0, -LRD2, \mbox{-SAT1}, and -SAT2. SAT1 fragments into three subcomponents: SAT1a, SAT1b, and SAT1c (see Fig. \ref{fig:SAT1b_fits}), although we treated it as a single galaxy here. 

We built a custom local lens model based on \cite{Gledhill2024}, incorporating information from three foreground galaxies close to the line of sight of our LRD system and using a new galaxy-galaxy lens that we discovered in one of these foreground sources (see Appendix \ref{app:lensing} for further details).
Our new lensing model yields a magnification of $\mu=1.26^{+0.04}_{-0.05}$ at the position of the LRD pair. We verified the result by replacing the external shear with the best cluster lens model from \citet{Gledhill2024}. The result ($\mu=1.302^{+0.007}_{-0.005}$) is consistent with our simplified model. The positions of our galaxies corrected for lensing are reported in Table \ref{tab:lensing_LRD}.

The five sources form a potential close group; the two brightest sources, LRD1 and LRD2, are at a projected distance (corrected for magnification and considering a scale of 5.4 kpc/arcsec) of only 3.27~kpc; LRD1 and SAT0 are separated by just 0.97~kpc; and LRD1 is separated from SAT1 and SAT2 by 6.35~kpc and 6.90~kpc, respectively.
Additionally, there is evidence of a tidal structure or an outflow connecting LRD1 to SAT2, as highlighted by the $F277W$ and $F410M$ smoothed maps in Fig.~\ref{fig:images}. At these $z$, these bands would be probing potential [OII]$\lambda\lambda$3726, 3729 and [OIII]$\lambda\lambda$4959, 5007 emission, respectively. The proximity of the objects in redshift and on the sky, as well as the hints of a physical connection between LRD1 and SAT2, are the first indications of the possible interacting nature of this galaxy group.

As also highlighted in Fig. \ref{fig:images}, there is another source to the north of our galaxy group (a possible SAT3) whose $F410M$ excess is indicative of $z\sim7$, compatible with the colors displayed by our satellite galaxies. It has a magnification $\mu = 1.35^{+0.05}_{-0.06}$ and is located $\sim18$ kpc away from the LRDs (corrected for lensing), and thus its physical connection to the galaxy group can be questioned. We report the photometry and physical properties of this source in Appendix \ref{app:SAT3}. 

\subsection{Revised photometry and LRD identification}\label{sec:PSFphotometry}

Given how close LRD1, LRD2, and SAT0 are, and the compactness of these objects, we went beyond the aperture photometry of Sect.~\ref{sec:data} and performed PSF photometry for these three sources. We retained the aperture photometry from the catalogs measured in 0.5\arcsec\, and 0.3\arcsec\, apertures for SAT1 and SAT2, respectively. Details regarding our measurements are included in Appendix~\ref{app:photometry}, and Table~\ref{tab:phot} reports the flux measurements for each source. Using the updated photometry, we recomputed photo-$z$s, arriving at the values listed in Table~\ref{tab:properties}.

To identify possible LRDs, we used the criteria for high-$z$ sources from \cite{Kokorev2024} and the criteria from \cite{Kocevski2024}. These and other LRD criteria are based on computing a set of colors along the SED, imposing a blue rest-frame UV continuum and a red optical continuum. They also include a compactness criterion and a screening for brown dwarfs. According to \cite{Kokorev2024}:
\begin{equation}
F150W - F200W < 0.8
\end{equation}
\begin{equation}
F277W - F356W > 0.6
\end{equation}
\begin{equation}
F277W - F444W > 0.7
\end{equation}
\begin{equation}
\texttt{compact} = f_{f444w}(0.4\arcsec)/f_{f444w}(0.2\arcsec) < 1.7
\end{equation}
\begin{equation}
\texttt{bd\_removal} = F115W - F200W > -0.5 
.\end{equation}

In this case, we used the fluxes measured within 0.3 and 0.5\arcsec\, apertures based on the PSF-matched images to compute the compactness. The color criteria were calculated using the PSF photometric values. According to \cite{Kocevski2024}:

\begin{equation}
\mathrm{S/N}_{F444W} > 12
\end{equation}
\begin{equation}
\beta_\mathrm{opt} > 0 
\end{equation}
\begin{equation}
-2.8 < \beta_{\mathrm{UV}} < -0.37
\end{equation}
\begin{equation}
r_{\mathrm{h}} < 1.5 r_{\mathrm{h,\,stars}}
\end{equation}
\begin{equation}
\beta_{F277W - F356W} > -1 
\end{equation}
\begin{equation}
\beta_{F277W - F410M} > -1
,\end{equation}where $r_\mathrm{h}$ stands for the half-light radius as measured in the $F444W$ band, $r_\mathrm{h, stars}$ denotes the stellar locus (as defined in their Fig. 4), S/N is the signal-to-noise ratio, and

$$\beta = \frac{0.4\,(\mathrm{m_1-m_2})}{\mathrm{log}(\lambda_2/\lambda_1)}-2.$$

To determine $\beta_{\mathrm{opt}}$ and $\beta_{\mathrm{opt}}$, we used the magnitudes measured in bands blueward and redward of 3645$\AA$ based on the $z$ of our galaxies, again corresponding to PSF photometry. The half-light radii were computed based on a Gaussian fit from the \texttt{RadialProfile} \texttt{photutils} method \citep[our pixel scale is 0.04\arcsec/pixel;][]{Bradley2023}. In Table \ref{tab:criteria} we include the values for the different criteria retrieved for the LRD pair and SAT0, as well as the compactness measured in every photometric band.

\begin{table*}[h]
\setlength{\tabcolsep}{1.1pt}
\renewcommand{\arraystretch}{1.5}
    \centering
    \footnotesize
    \caption{Values for the criteria established in \citet{Kokorev2024} and \citet{Kocevski2024}, defined in Sect. \ref{sec:PSFphotometry}, obtained for the LRD pair and SAT0.}
    \begin{tabular}{c|c|c|c|c|c||c|c|c|c|c|c||c|c|c|c|c|c|c|c|c}
    Source& \multicolumn{5}{c||}{Kokorev+24}&\multicolumn{6}{c||}{Kocevski+24}&\multicolumn{9}{c}{Compactness}\\ 
    \hline
    & (1)&(2)&(3)&(4)&(5)&(6)&(7)&(8)&(9)&(10)&(11)&$F606W$&$F814W$&$F090W$&$F115W$&$F150W$&$F200W$&$F277W$&$F356W$&$F410M$\\&(mag)&(mag)&(mag)&&(mag)&&&&(pix)&&&&&&&&&&&\\
    \hline\hline

    LRD1& 0.02 & 1.57& 1.90&1.24 &0.20& 252&1.73&$-$1.65&1.03&3.78&2.72&-&2.08&1.58&1.68&1.59&1.55&1.51&1.35&1.35\\

    LRD2& 0.08 & 1.49& 1.73&1.20 &0.62&59 &1.38&$-0.98$&1.72&3.49&2.38&-& 0.24&-&0.90&1.29&1.11&1.22&1.21&1.19\\

    SAT0& $-0.17$ & 1.11& 0.87&\textbf{2.03} &$-0.06$&13&\textbf{$-$0.30}&$-2.10$&2.41&2.08&1.48&-&1.10&1.40&1.48&1.50&1.54&1.61&1.80&1.81\\
\hline
    \end{tabular}
    \label{tab:criteria}
    
    \tablefoot{Compactness refers to $f$(0.5\arcsec)/$f$(0.3\arcsec) in the corresponding band. Bold values highlight those criteria that are not satisfied. Absent values correspond to measurements in which fluxes are negative. Note that measurements of compactness and half-light radius of SAT0 are highly susceptible to contamination from LRD1. The LRDs display a compactness $<1.7$ also in the rest of the bands. This value is larger in $F814W$ in the case of LRD1, but measurements in that band are quite noisy (S/N $\sim$ 0.8 from PSF photometry).}
\end{table*}

LRD1 and LRD2 pass these screenings, while SAT0 satisfies most of the color criteria but its LRD nature is more uncertain as its proximity to LRD1 involves a certain contamination, even for the PSF photometry. This contamination also makes it difficult to check its compactness. In addition, as highlighted by the red-green-blue (RGB) images in Fig. \ref{fig:images}, SAT0 shows different colors from those displayed by the LRDs, more compatible with SAT1 and SAT2. We cannot make any further conclusions about the nature of this object with our data and will treat it as a satellite source.

LRD1 reaches $\sim25$ mag in the rest-frame optical and $\sim27$ mag in the UV, which places it among the brightest LRDs known in the legacy fields from \citet{Kocevski2024}. LRD2 is fainter, reaching $\sim26$ mag in the optical and $\sim28$ mag in the UV, comparable to typical LRDs. The satellites are also fainter, and SAT0 is the brightest among them. It reaches $\sim25.5 - 26.5$ mag in the optical and $27$ mag in the UV. SAT1 and SAT2 display average values of 28 and 28.5 mag, respectively.

\section{SED fitting}
\label{sec:sed}

We fitted the photometry of LRD1, LRD2, and the satellites using \texttt{Dense Basis} (\citealt{Iyer2017}, \citealt{Iyer2019}) and \texttt{Bagpipes} \citep{Carnall2018}. One of the primary advantages of \texttt{Dense Basis} is that it utilizes nonparametric SFHs, which work better than parametric models capturing the variety of physical events (e.g., bursts, rejuvenation, sudden quenching, or maximally old star formation) that shape the complex SFHs of high-$z$ galaxies (\citealt{Simha2014}, \citealt{Leja2019}). Stellar population synthesis models are incorporated into the code through the Flexible Stellar Population Synthesis (\texttt{FSPS}; \citealt{Conroy2010}) \texttt{Python} module. It allows for the inclusion of the infrared emission of an AGN's dusty torus based on the \citealt{Nenkova2008} templates, but the emission from the accretion disk in the UV-to-optical is not considered. Consequently, \texttt{Dense Basis} can only be effectively used to fit the emission due to stars at these wavelengths.

On the other hand, \texttt{Bagpipes} allows the UV and optical emission due to an AGN accretion disk to be modeled using a double power-law model. A parameterization for the SFH is normally selected in this code, although the code also allows for nonparametric SFHs in a different way than \texttt{Dense Basis}. It uses a series of piecewise constant functions in lookback time, requiring a number of bins, a prior distribution, and an \texttt{alpha} parameter that controls the correlation between the t$_{\mathrm{x}}$ (the age at which the galaxy formed x\% of its $M_\star$) and the SFH \citep{Leja2019}. \texttt{Dense Basis} does not require these time bins and we can choose a set of custom values of \texttt{alpha} for each parameter. 

We opted to use the \texttt{Dense Basis} code to fit the stellar emission and recover the SFHs of these galaxies and explored the AGN scenario using \texttt{Bagpipes}, and a combination of the two codes. More details regarding the \texttt{Dense Basis} and \texttt{Bagpipes} methods and configurations can be found in Appendices \ref{app:DB} and \ref{app:bagpipes}, respectively. The physical parameters obtained for each source are reported in Table \ref{tab:properties}.

The interplay between the peak of the Ly$\alpha$ emission and the damping wing absorption plays a significant role at this redshift, as mentioned in Sect. \ref{sec:data}. Since \texttt{Dense Basis} and \texttt{Bagpipes} use the default IGM attenuation curve, and the attenuation is in any case stochastic, we did not consider filters close to this feature in the SED fitting, namely $F814W$, $F090W$, and $F115W$. Our effective photometry comprises six filters (five broad bands and one medium band) spanning from $F150W$ to $F444W$. We are aware that our data are highly susceptible to overfitting and thus results are highly model-dependent. However, this study aims to provide just initial insights into the nature of these objects and their SFHs.

When using a broad prior for the redshift in these codes, both \texttt{Dense Basis} and \texttt{Bagpipes} favor a $z \sim 7$ solution for the LRD pair (using the six filters mentioned above, not considering the filters close to the Ly$\alpha$ damping wing). These results are in line with the photo-$z$s reported by \texttt{EAzY} when using the \cite{Killi2024} template. However, the models used by SED fitting codes such as \texttt{Dense Basis} and \texttt{Bagpipes} still need to be fine-tuned to account for the physical conditions ruling at these early epochs. Furthermore, LRDs are very peculiar objects that currently present a challenge to SED fitting codes. Considering that the satellites' photo-$z$s are less uncertain since these are normal star-forming galaxies (SFGs), we can use these values as the redshift of the galaxy group (i.e., $z\sim6.8)$. To do so, we restricted the redshift prior to $6.75 < z < 6.85$ in \texttt{Dense Basis} and \texttt{Bagpipes}.

\begin{figure*}[htp]
\centering
    \begin{subfigure}{0.9\textwidth}
    \centering
    \includegraphics[width=\textwidth]{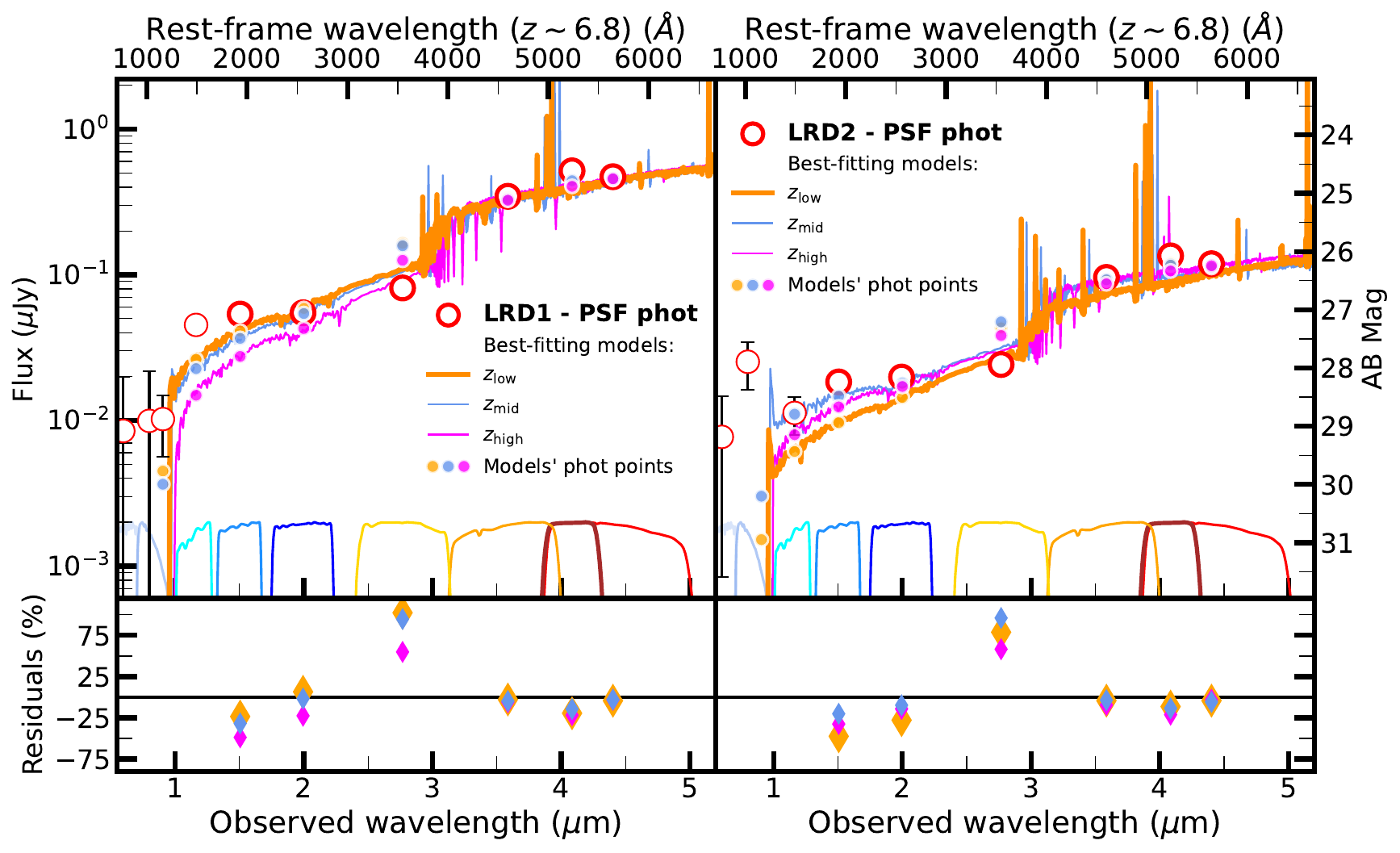} 
    \end{subfigure}
    \hfill
    
    \begin{subfigure}{\textwidth}
    \centering
    \includegraphics[width=\textwidth]{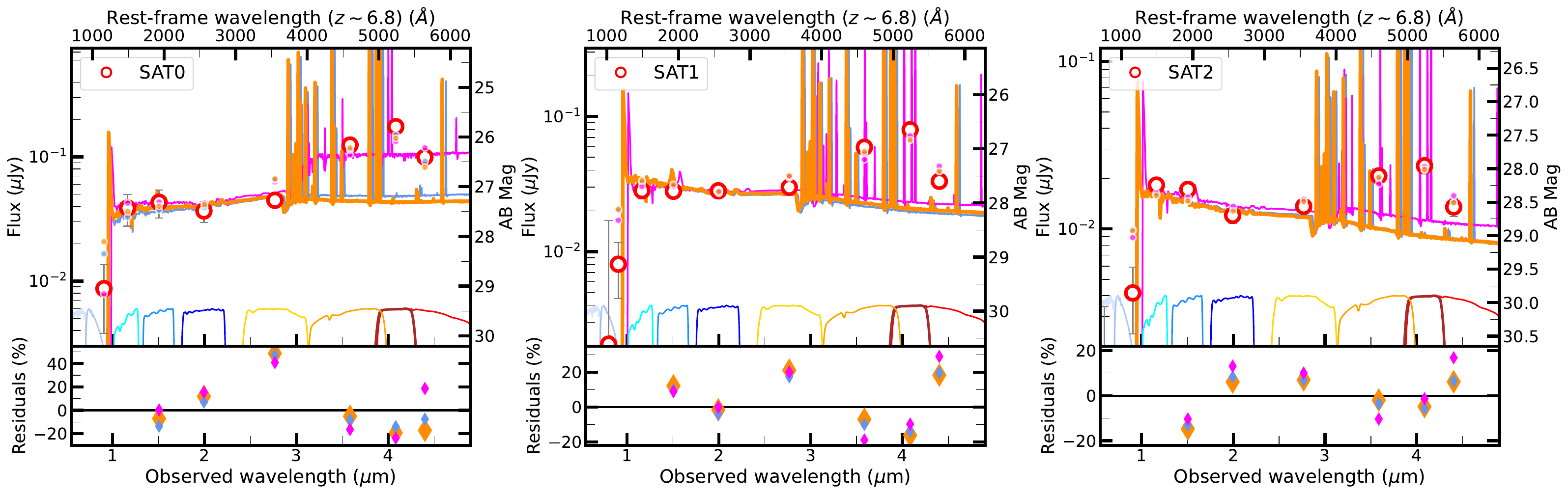}
    \end{subfigure}
    
    \caption{(a): \texttt{Dense Basis} stellar fits based on the LRD1 (left) and LRD2 (right) PSF photometry (red open circles). Data points excluded from the fits are shown using thinner lines. Solid orange, blue, and fuchsia lines depict the fits for the low- (thicker line, corresponding to the photo-$z$ of the galaxy group), mid-, and high-$z$ intervals. The corresponding model photometric points are shown with filled circles. (b): \texttt{Dense Basis} stellar fits for the satellite galaxies based on PSF (SAT0) and aperture (SAT1, SAT2) photometry, following the same color code as above. We include the HST and NIRCam transmission curves, as well as the residuals of the fits.}
    \label{fig:DB_LRDs_fits}
\end{figure*}

In addition, given the available data and the limitations and biases of our SED fitting codes, we opted to be cautious and explored a more conservative approach, considering small variations to this prior. These variations also account for the possibility that the LRDs and the satellites were not located at the same $z$, as well as the random IGM and/or CGM attenuation variations along the line of sight, which could potentially lead to small variations in our galaxies' photo-$z$s. To deal with this, we added two more redshift ranges, restricting the priors to $6.75 < z < 6.95$ (allowing for a slightly higher redshift), and $6.95 < z < 7.15$ (to explore the $z\sim7$ case). We refer to these intervals as the low- ($6.75 < z < 6.85$), mid- ($6.75 < z < 6.95$), and high-$z$ ($6.95 < z < 7.15$) cases hereafter, recalling that the low-$z$ interval is the one centered around the actual photo-$z$ of this potential galaxy group.

\subsection{Stellar fits from \texttt{Dense Basis}}
\label{sec:DB}  

\begin{figure*}[htp]
    \centering    
    \includegraphics[width=\linewidth]{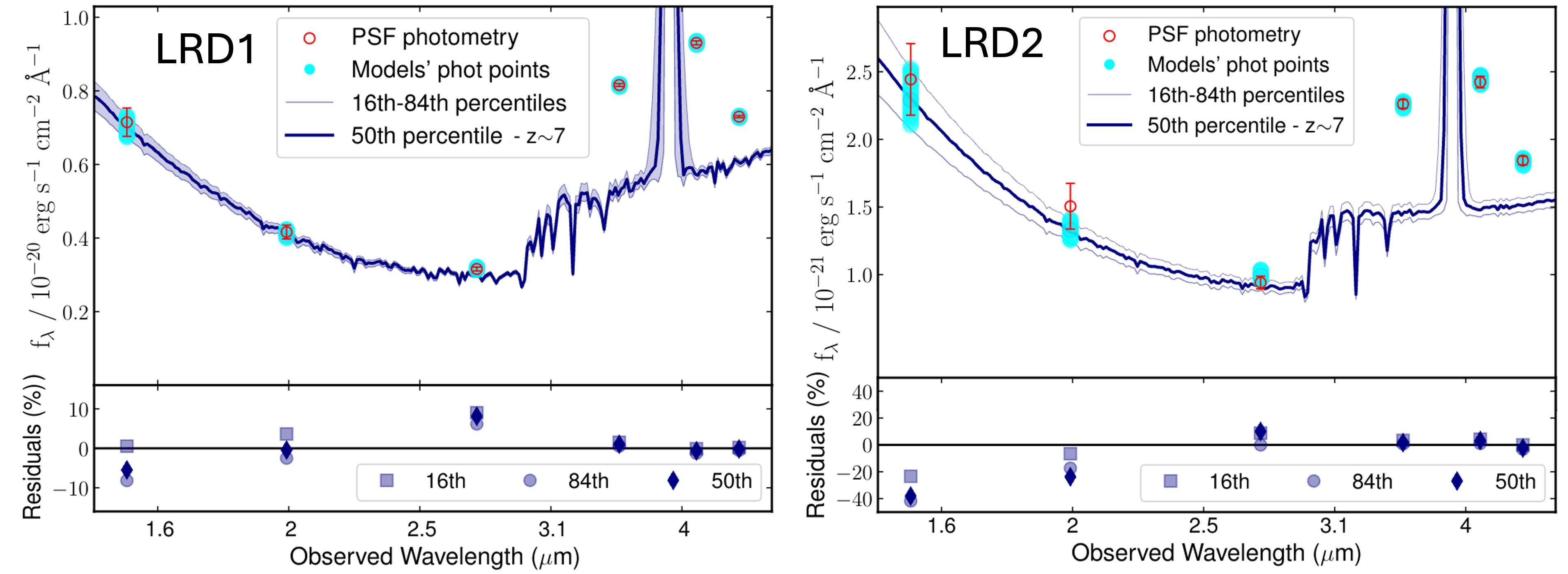}
    \caption{\texttt{Bagpipes} best fits (stars + AGN) for LRD1 (left) and LRD2 (right) based on PSF photometry (red open circles), corresponding to the high-$z$ solution obtained using a broad $z$ prior. Cyan circles depict the synthetic photometric points associated with the best-fitting model, shown as a thick blue line (50th percentile). The limits of the shaded blue region represent the 16th and 84th percentiles. The residuals for the models, corresponding to the 16th (squares), 50th (diamonds), and 84th (circles) percentiles, are also included underneath each panel.
    } 
\label{fig:bagpipes_lrd1_highz}
\end{figure*}

We assumed wide ranges for the priors of the stellar mass and SFR, constraining the $M_\star$ between \mbox{$7<\mathrm{log}\, M_\star/M_\odot<12$} and the SFR between \mbox{$-1<\mathrm{log\, SFR} [M_\odot/\mathrm{yr}]<3$}, and setting a flat specific SFR. For parameters such as dust attenuations or metallicities, the default \texttt{Dense Basis} configuration was used, which considers a \cite{Calzetti2000} dust attenuation law. Additional details regarding our setting can be found in Appendix \ref{app:DB}.

The left panel in the first row of Fig. \ref{fig:DB_LRDs_fits} shows the stellar best-fitting models from \texttt{Dense Basis} for LRD1.
These models correspond to a dusty (A(V) $\sim1.0 - 1.2$ mag) and massive (\Mstars~$\sim10^{10.4-10.5} M_\odot$) Balmer break galaxy. Residuals show that the solutions that best fit the $F150W$ and $F200W$ points (e.g., the low- and mid-$z$ cases, with $<25$\% residuals) deviate more from the $F277W$ point ($>75\%$). LRD1 is thus poorly fitted using only stars, especially in the rest-frame UV, where the code cannot well reproduce the flat slope and provides a steeper model with a high A(V) (see Appendix \ref{app:DB} for more details on priors and tests). All these factors indicate a potential additional component, responsible for the UV flux excess. 

LRD2's stellar fit is displayed in the right panel, first row, of Fig. \ref{fig:DB_LRDs_fits}. All of the solutions indicate a massive (\Mstars~$\sim10^{9.7-9.8} M_\odot$) Balmer break galaxy, subject to an uncertain degree of attenuation (A(V) ${\sim0.62 - 1.26}$ mag). Residuals highlight the difficulty of fitting the break as defined by the $F277W$ point while accounting for the UV flux. However, the high-$z$ case seems to provide a better balance between the two tasks, fitting the UV and the break with residuals $\lesssim 25$\% in $F150W$ and $\sim50$\% in $F277W$, suggesting a potential stellar origin for this source. 

Previous studies have also found Balmer break LRDs spanning $M_\star \sim 10^{9-11}\, M_\odot$ (see \citealt{Gentile2024} at $z=5.051$ and 6.7; \citealt{Ma2024} at $z=7.04$; \citealt{Wang2024} at $6.7<z<8.4$; \citealt{Williams2024} at $5.5<z<7.7$; \citealt{Labbe2024} at $z=4.47$). In fact, \citet{Wang2024} provide a sample of spectroscopically confirmed Balmer break LRDs at the same $z_{\mathrm{spec}}$ of our sources, suggesting that such Balmer breaks are possible in LRDs at this epoch.
The number of reported LRDs with a Balmer break is probably too low as some might have been discarded in works whose screenings excluded galaxies that show strong emission lines and/or strong breaks to ensure a rising red continuum (e.g., \citealt{Labbe2023}, \citealt{Greene2024}, \citealt{Kocevski2024}, \citealt{Kokorev2024}). 

However, it is also important to remember that emission lines can mimic Balmer breaks, which would lead to overestimated values of $M_\star$ \citep{Desprez2024}, as in a fraction of objects from \citet{Labbe2023}. \citet{Labbe2023} was based on the analysis of NIRCam broad-band photometry, while we benefit from an emission excess in a medium-band filter, $F410M$, likely produced by [OIII]$\lambda\lambda 4959, 5007$+H$\beta$, which also allows us to better constrain the photo-$z$s of our galaxies. Still, we cannot rule out the possibility that these breaks are partially or totally induced by emission lines. Further spectroscopic confirmation of this feature is required. Additional discussion regarding potential contamination due to emission lines can be found in Appendix \ref{app:DB}.

Panels in the second row in Fig. \ref{fig:DB_LRDs_fits} depict the stellar fits for the satellites. SAT0 is potentially a SFG. It can also be fitted with a Balmer break at $z\sim7$, although the colors of this source according to Fig.~\ref{fig:images} favor the SFG scenario. All the fits suggest a massive object subject to a low dust obscuration, with $M_\star \sim 10^{9.2-9.4} M_\odot$ and ${\mathrm{A(V)}\, \sim 0.1 - 0.3}$ mag. SAT1 and SAT2 are lower-mass $\sim10^{8.0}\,M_\odot$ SFGs with low attenuations (A(V) $<0.3$ mag) at the three redshift intervals. Residuals for SAT1 and SAT2 are small, $\lesssim20$\%, highlighting that the source of the high residuals in the LRD fits is not our methodology but rather the unusual nature of these targets.

\subsection{AGN modeling with \texttt{Bagpipes}}
\label{sec:bagpipes}

The results obtained with \texttt{Dense Basis} for the LRDs, especially LRD1, point to a UV excess that cannot be fitted only with stars with current models. Several recent works suggest that this emission could result from extremely low dust attenuation conditions in which dust is not destroyed but pushed out by kiloparsec-scale outflows (the attenuation-free model; \citealt{Ferrara2023}, \citealt{Ziparo2023}, \citealt{Ferrara2024a,Ferrara2024b}). A top-heavy IMF (\citealt{Trinca2024}, \citealt{Hutter2024}), a higher star formation efficiency (\citealt{Dekel2023}, \citealt{Li2024}), or star formation variability (\citealt{Mirocha2023}, \citealt{Pallottini2023}) are also being explored as possible drivers of these blue colors. On the other hand, this emission could be powered by a nonstellar source, such as an AGN. In this section we explore this last possibility using \texttt{Bagpipes} (see also Appendix \ref{app:bagpipes}), keeping in mind that this is just only one of the different potential scenarios that could explain this excess in our LRDs. 

We fitted the photometric points of the two LRDs using a composite model consisting of a stellar population plus an AGN employing the Multinest sampling algorithm. In the case of LRD1, \texttt{Bagpipes} chooses a $z\sim7$ solution, as previously mentioned. For the low-$z$ prior (the photo-$z$ of the galaxy group, $z\sim6.8$), the fit does not converge, probably due to limits in the code (see Sect. \ref{sec:sed}). We explored the other two redshift cases to provide some hints about the possible AGN nature of this LRD.

\begin{figure*}[htp]
\centering
    \centering
    \includegraphics[width=\textwidth]{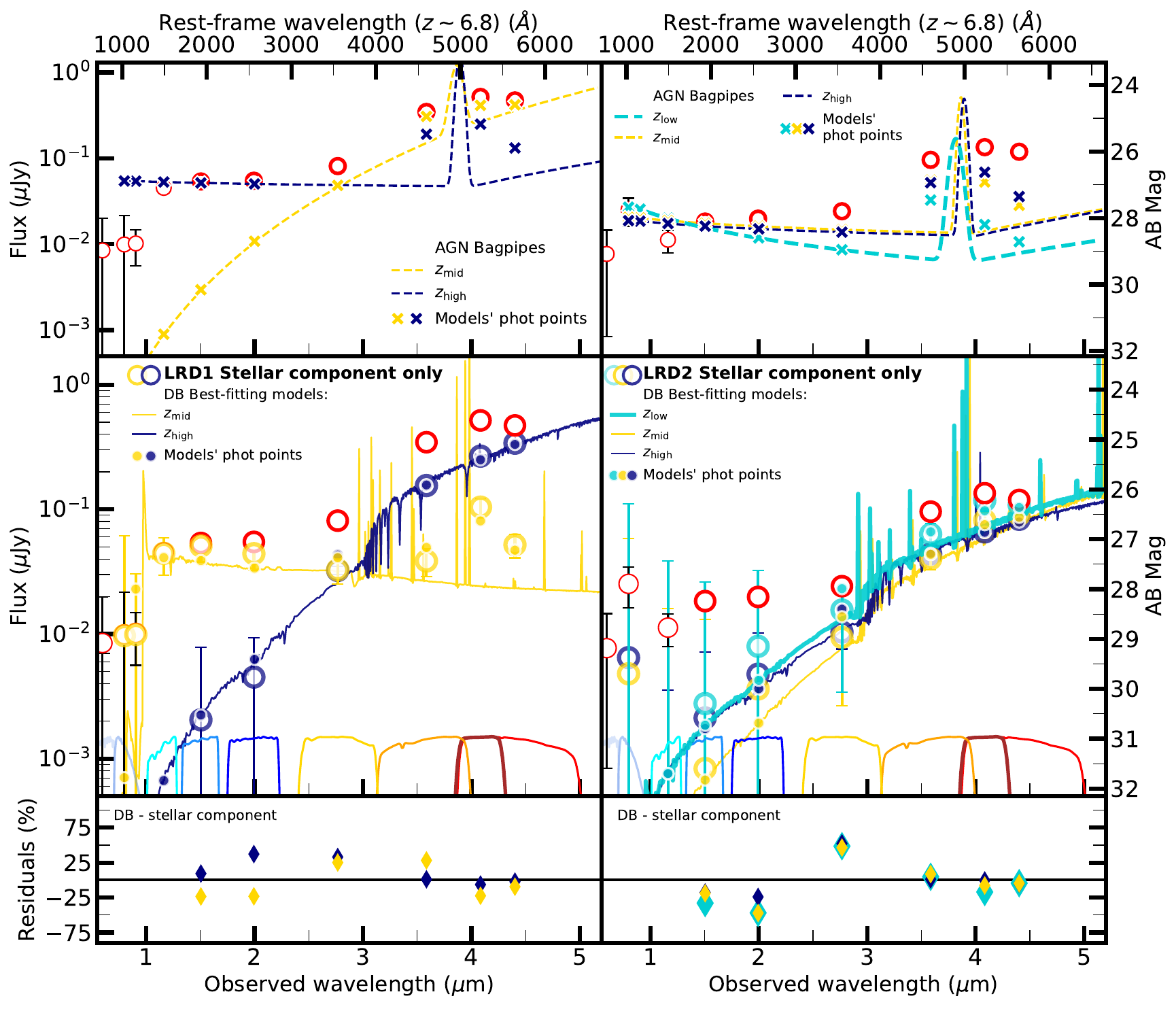} 
    \caption{Top panels: Potential AGN fits (dashed cyan for the low-$z$, yellow for the mid-$z$, navy for the high-$z$ cases) for LRD1 (left) and LRD2 (right) derived using \texttt{Bagpipes} (Sect. \ref{sec:bagpipes}). The low-$z$ case is not represented in the LRD1 panel (left) because it did not converge in \texttt{Bagpipes}. Crosses depict their photometric points. Middle panels: \texttt{Dense Basis} fits for the stellar component of LRD1 (left) and LRD2 (right) obtained by subtracting the AGN photometry derived with \texttt{Bagpipes} from the PSF photometry (Sect. \ref{sec:stellar_component}). The LRD PSF photometry is shown as red open circles. Photometric points from the stellar component are shown as cyan, yellow, and navy open circles for the low-, mid-, and high-$z$ cases, respectively. Best-fit models are displayed as solid lines in the same color. The low-$z$ case (thicker line) corresponds to the photo-$z$ of the galaxy group. Photometric points for the models are depicted as filled circles. HST and NIRCam transmission curves are also included. Bottom panels: Residuals of the fits.}
    \label{fig:DB_BP_LRDs_fits}
\end{figure*}

The best-fitting model from \texttt{Bagpipes} (stars+AGN) for LRD1 ($z\sim7$) is presented in Fig. \ref{fig:bagpipes_lrd1_highz}. Its isolated AGN photometric component is shown in the top panel of Fig. \ref{fig:DB_BP_LRDs_fits}. The AGN continuum is modeled by an unobscured broken power-law dominating the UV, which has been found to reproduce well the average observed Sloan Digital Sky Survey (SDSS) quasar spectra (e.g., \citealt{VanderBerk2001}, \citealt{Temple2021}). This model has also been proposed for LRDs at high-$z$ (e.g., \citealt{Kocevski2023}, \citealt{Tripodi2024}, \citealt{Zhang2024}). On the other hand, a broad H$\beta$ emission line and a reddened stellar population dominate the optical SED. 

The solution for the mid-$z$ case does not correspond to a minimum in the $z$ parameter space. Nevertheless, we show the results for this mid-$z$ AGN as it describes one of the scenarios that have been proposed to explain the nature of LRDs. This solution (Fig. \ref{fig:LRD1_bagpipes_zint} and the top panel in Fig. \ref{fig:DB_BP_LRDs_fits}) resembles a dust-obscured AGN that dominates the emission in the optical, similarly to the preferred scenario illustrated in \citet{Greene2024}, \citet{Kocevski2024}, \citet{Wang2024}, and \citet{Ma2024}. The rest-frame UV would be attributed to scattered light from the AGN (as in \citealt{Perez-Gonzalez2024} for their NIRCam analysis) and/or stars (as in \citealt{Greene2024}). In LRD1, the AGN at the high-$z$ (mid-$z$) range would produce 96(5)\%, 92(20)\%, 60(60)\%, 55(89)\%, 48(80)\%, and 28(89)\% of the flux in $F150W$, $F200W$, $F277W$, $F356W$, $F410M$, and $F444W$, respectively.

\texttt{Bagpipes} also favors a $z=7$ for LRD2, but it can find a solution for each of the three redshift intervals. All of these models correspond to the unobscured AGN scenario; see Figs. \ref{fig:bagpipes_lrd1_highz} ($z\sim7)$ and \ref{fig:LRD2_bagpipes} (low- and mid-$z$ cases) for the stars+AGN fits and the top panel in Fig. \ref{fig:DB_BP_LRDs_fits} for the isolated AGN components. In LRD2, the AGN would be responsible for 88\%, 76\%, 57\%, 54\%, 51\%, and 30\% of the emission in the NIRCam filters in the high-$z$ case;  95\%, 81\%, 61\%, 58\%, 39\%, and 24\% in the mid-$z$ case; and 85\%, 60\%, 37\%, 34\%, 13\%, and 9\% in the low-$z$ case.

\subsection{\texttt{Dense Basis} + \texttt{Bagpipes}. An approach to isolating the stellar component}
\label{sec:stellar_component}

As shown in Fig.~\ref{fig:bagpipes_lrd1_highz}, composite models from \texttt{Bagpipes} also favor a Balmer break over a star-forming solution. Residuals for LRD1 are much lower ($\leq 10$\%) compared to the \texttt{Dense Basis} fit, also in $F277W$, providing further support to the hybrid model hypothesis. 
The presence of an AGN in LRD2 is more ambiguous, given \texttt{Bagpipes} residuals, but it cannot be ruled out. It may differ from the AGN models considered here or there could be another physical component that we have not identified yet.

For the LRD1 host, \texttt{Bagpipes} reports $M_\star\sim 10^{11.11} \,M_\odot$ and A(V) = 2.29 mag (at $z\sim7$) while for LRD2 these values are $M_\star = 10^{10.0-10.6}\, M_\odot $ and A(V) = $2.34-2.46$ mag, considering the three $z$ cases. When the AGN emission is taken into account, dustier and thus more massive hosts are revealed, mostly responsible for the rest-frame optical emission. Compared to other LRDs in the literature, these galaxies show average to low values of dust attenuation, and larger $M_\star$ (e.g., \citealt{Kocevski2024} sources display an average of A(V) $\sim2.8$ mag and $M_\star \sim 10^9\, M_\odot$).
However, we warn that a $M_\star\sim10^{11}\,M_\odot$ is unlikely at such early times (although see \citealt{Gentile2024}) and would challenge the current cosmological framework if they existed in large numbers at these redshifts (\citealt{Menci2022}, \citealt{Boylan2023}, \citealt{Desprez2024}). $M_\star$ is a rather uncertain parameter in LRDs (e.g., \citealt{Tripodi2024}, \citealt{Wang2024}), degenerate with attenuation and AGN contribution, and should be interpreted with extreme caution.

Being aware that \texttt{Dense Basis} and \texttt{Bagpipes} work differently and that each one is based on a different set of assumptions, in this section we present an approach to combining their power and obtaining an estimate of the emission due solely to the stellar component in the LRDs. We then fit this estimated stellar component with \texttt{Dense Basis} and its nonparametric SFH method. We estimate the stars-only photometric component by subtracting the photometric points of the \texttt{Bagpipes} AGN model from the observed LRD PSF photometry. These results should only be interpreted qualitatively and as a sanity check to those obtained with \texttt{Bagpipes}.

In the bottom left panel of Fig. \ref{fig:DB_BP_LRDs_fits}, we show the LRD1 photometry corresponding to the stellar component, as well as the \texttt{Dense Basis} stellar best-fitting solutions. 
If a dust-free AGN existed in this galaxy (the high-$z$ case AGN), the stellar component would exhibit a high level of attenuation (A(V) $\sim$ 2.8 mag) and a very high $M_\star$ ($\sim10^{11.2} M_\odot$). On the other hand, if the AGN were highly obscured and dominated the optical emission (the mid-$z$ case), the stellar component would be dust-free (A(V) = 0.07~mag) and linked to a low-mass galaxy of $\sim10^{8.2} M_\odot$, resembling source MSAID38108 from \citet{Chen2024} or COS-6696 from \citet{Akins2024}. Even though this method involves significant caveats, we could recover a very similar stellar component to that fitted by \texttt{Bagpipes}, with residuals $<25$\% in all the bands, including $F277W$.

The stellar component of LRD2 is displayed in the bottom right panel of Fig. \ref{fig:DB_BP_LRDs_fits}. It indicates the presence of highly attenuated stellar populations (A(V) = $1.88 - 2.73$ mag) with $M_\star>10^{10} M_\odot$. The difference in the residuals between the original photometry and the stellar component is not so evident and the $F277W$ filter continues to present a challenge in the fitting ($\sim50$\% residual), in line with \texttt{Bagpipes} results. 

In Table \ref{tab:summary} we provide a qualitative description of the different possibilities discussed along Sects. \ref{sec:DB}, \ref{sec:bagpipes}, and \ref{sec:stellar_component} for this potential galaxy group, based on \texttt{Dense Basis} and \texttt{Bagpipes} SED fitting.

\subsection{Star formation histories}
\label{sec:sfhs}

\begin{figure*}[htp]
\centering
    \begin{subfigure}{0.9\textwidth}
    \centering
    \includegraphics[width=\textwidth]{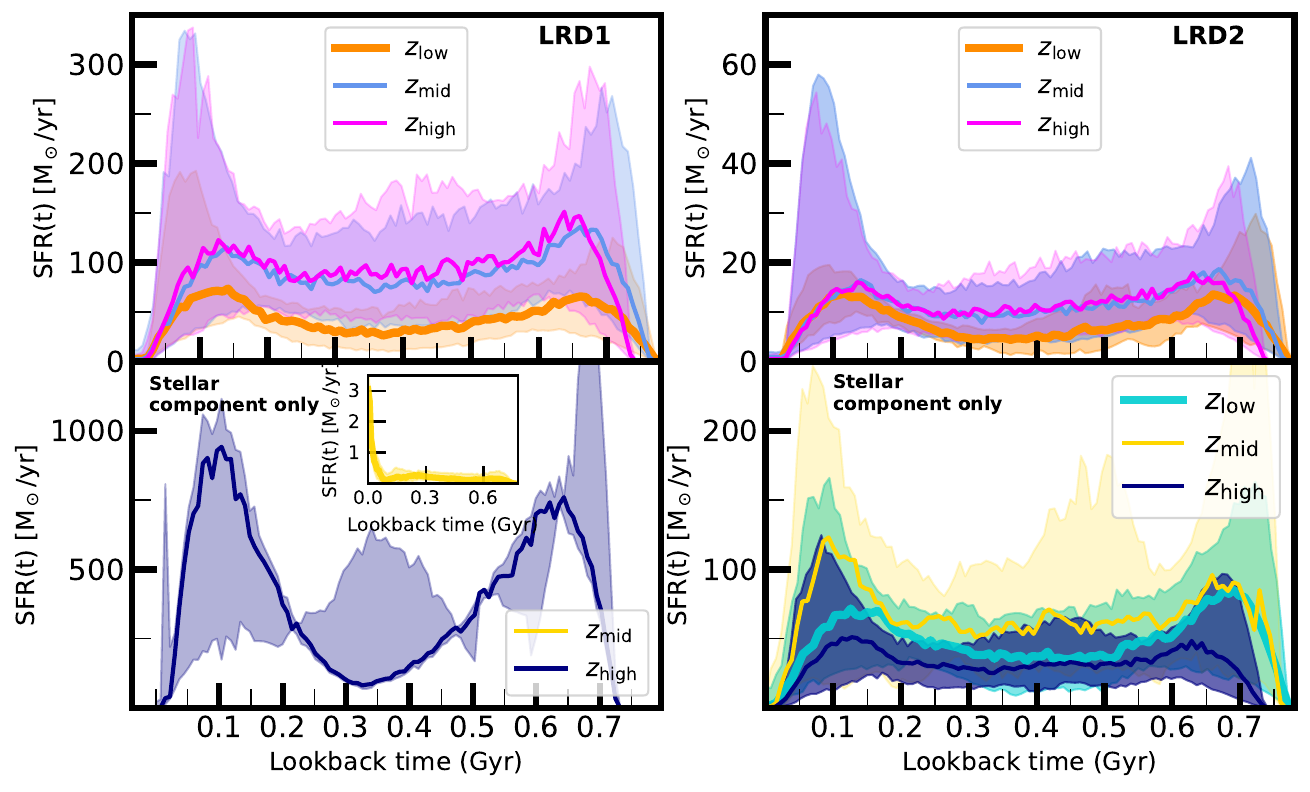} 
    \end{subfigure}
    \hfill
    \begin{subfigure}{\textwidth}
    \centering
    \includegraphics[width=\textwidth]{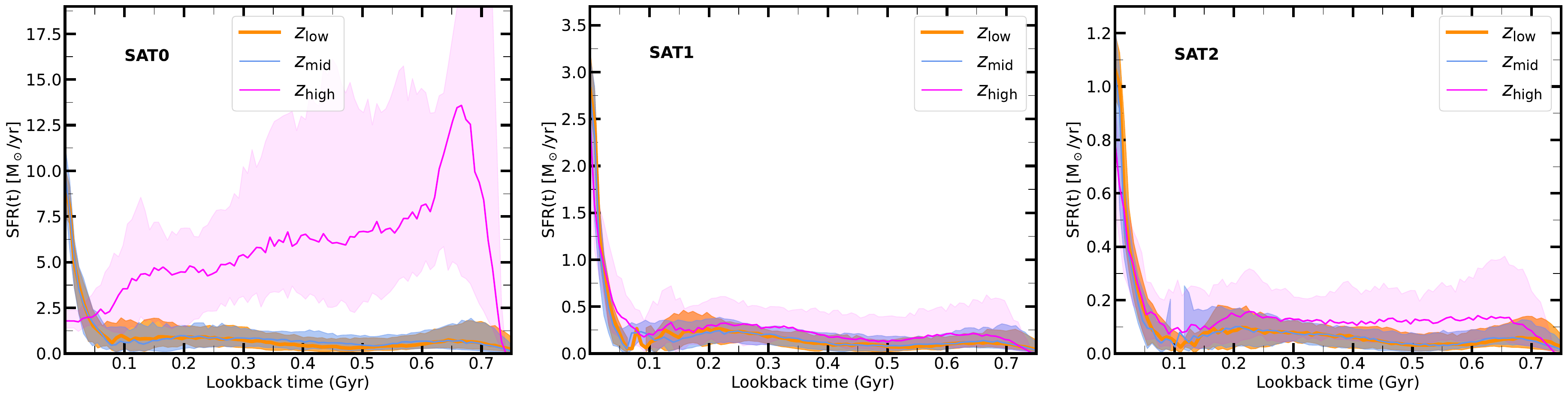}
    \end{subfigure}
    \caption{Top panels: \texttt{Dense Basis} SFHs based on the LRD1 (left) and LRD2 (right) PSF photometry. Solid lines show the 50th percentile and shaded regions the 16th–84th percentiles, color-coded by redshift. The thicker line depicts the low-$z$ case, which corresponds to the photo-$z$ of the galaxy group. Middle panels: \texttt{Dense Basis} SFHs derived from the stellar component only (obtained by subtracting the AGN photometry from the PSF photometry; Sect. \ref{sec:stellar_component}), color-coded by redshift. The inset in the left panel shows a zoomed-in view of the LRD1 mid-$z$ SFH. Bottom panels (from left to right): SFHs derived for SAT0, SAT1, and SAT2, color-coded by redshift.}
    \label{fig:DB_LRDs_SFH}
\end{figure*}

The SFHs of the LRD pair are presented in the first row in Fig.~\ref{fig:DB_LRDs_SFH}. Results derived by \texttt{Dense Basis} from PSF photometry (top panels; see Sect. \ref{sec:DB}) reveal a relatively smooth trajectory for both LRDs, with a fairly constant and uninterrupted star formation period followed by recent quenching. LRD1 formed on average $\sim100\, M_\odot$/yr over the last $\sim$700~Myr, while the star formation in LRD2 was comparatively modest ($\sim15\, M_\odot$/yr). It is possible that the SFHs of both LRDs are reflecting a recent encounter that took place in the last 200 Myr. In this period, according to the 50th percentile, both LRDs appear to have undergone synchronized bursts of star formation, peaking $100-150$ Myr ago and followed by quenching. However, the envelopes of both distributions are also consistent with a scenario in which the star formation peak of one of the galaxies could be delayed with respect to the other. They both look quenched at $z\sim7$, yet the quenching would not have been synchronous in this case. We focus the discussion on these previous possibilities, although it is worth noting that the envelopes of the distributions could also allow for a direct quenching without an earlier burst of star formation. 

This quenching is compatible with the prevailing idea that Balmer breaks are associated with evolved stellar populations, although these breaks could also originate from dense gas absorption near the AGN (\citealt{Inayoshi2024}; see also \citealt{deGraaf2025}, \citealt{Ji2025}, \citealt{Naidu2025}, and \citealt{Rusakov2025}).

The SFHs obtained based on the stellar component of the stars+AGN scenario (see Sect.~\ref{sec:stellar_component} and the bottom panels in the first row in Fig. \ref{fig:DB_LRDs_SFH}) yield the same results in the case of LRD2, although with a higher average SFR. For LRD1, the high-$z$ case (a dust-free AGN) provides a burstier, less smooth SFH compared to our previous results. It also reflects a recent burst of star formation, reaching 900 $M_\odot/$yr. According to the mid-$z$ solution, which resembles the obscured AGN, LRD1 formed stars at a very low rate and is currently undergoing a burst of star formation. The mid-$z$ AGN is not a good solution for LRD1 (see Sect. \ref{sec:bagpipes}) and it is only included to illustrate a possible option to describe the LRDs' SEDs.
Further discussion for LRD1 will be based on the SFH provided by the high-$z$ case, keeping in mind that the photo-$z$ of the galaxy group, based on the potential satellites, is slightly lower, but our codes cannot find a good solution if we do not allow a $z\sim7$ fit. Alternative scenarios could also reproduce our observables as well (see Sect.~\ref{sec:bagpipes}). 

The SFHs of the satellites are displayed in the second row in Fig. \ref{fig:DB_LRDs_SFH}. SAT0 shows two potential evolutionary pathways, corresponding to the star-forming and the Balmer break solutions. The SFHs of SAT1 and SAT2 are consistent with recent bursts of star formation that started $\sim100$ Myr ago, coincident with the bursts observed in the LRDs.

\begin{figure*}[htp]
    \centering
    \includegraphics[width=0.9\linewidth]{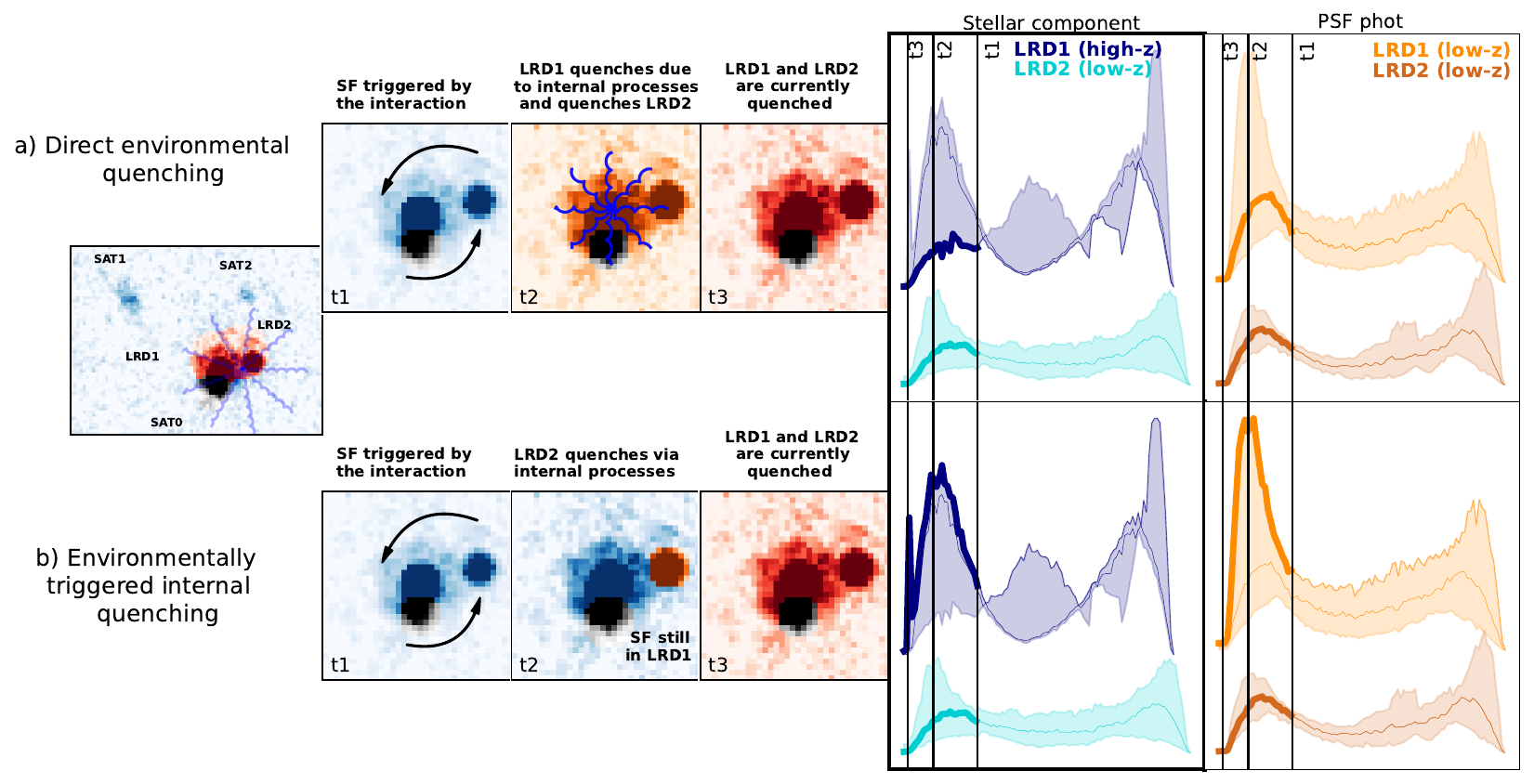}
    \caption{Potential scenarios for the LRDs' SFHs. Red colors in the image cutouts indicate quenching, orange a quenching in progress, and blue star formation. SAT0 is depicted in black due to its uncertain nature. The two rows show three time steps within the last $\sim200$ Myr, marked in the SFH plots as vertical lines. Case (a) corresponds to direct environmental quenching, and case (b) shows quenching due to internal processes, triggered by the environment. LRD1 is treated as the AGN host, but the same would apply if LRD2 or both LRDs were AGN hosts. In the SFH plots (last two columns), thicker lines highlight the curves used as a reference for each case. We show the SFHs obtained with \texttt{Dense Basis} for the stellar component only (highlighted with thicker borders), derived by subtracting the AGN photometric points from \texttt{Bagpipes} from the PSF photometry (Sect. \ref{sec:stellar_component}). The SFHs based on the untouched PSF photometry (Sect. \ref{sec:DB}) are depicted in the last column. In the panels showing the SFHs based on the stellar component, we represent the LRD1 results for the high-$z$ case in navy (see Sect. \ref{sec:bagpipes}) and the LRD2 results for the low-$z$ case (centered at the photo-$z$ of the galaxy group) in cyan, following the Fig. \ref{fig:DB_LRDs_SFH} color code. The low-$z$ SFHs are represented in orange (LRD1) and brown (LRD2) in the last column, which shows the SFHs based on the untouched PSF photometry, following the Fig. \ref{fig:DB_LRDs_SFH} color code. The scaling of the SFHs is arbitrary in this cartoon and thus not indicated.}
    \label{fig:scenarios}
\end{figure*}

\section{Discussion}
\label{sec:discussion}

\label{sec:discussion_SFHs}

The caveats involved in our strategy to investigate the nature of this LRD pair are numerous and difficult to tackle with our data. Spectroscopic redshifts, better sampling of these potential Balmer breaks, and other AGN diagnostics are required. Thus, rather than trying to comprehensively ``solve'' this system, in this work we aim to establish the basis for future research, providing some hypotheses about the physical processes taking place inside and outside these rich high-$z$ laboratories. 

The main novelty of this study is the observation of a pair of LRDs in close proximity (see also \citealt{Tanaka2024}), embedded in a potential system of satellites at an early stage of the Universe's life. If these LRDs were Balmer break galaxies, as reported by \texttt{Dense Basis} and \texttt{Bagpipes}, this pair would have gone through recent bursts of star formation, followed by quenching, during the last 200 Myr. The satellites are experiencing bursts of star formation that started 100 Myr ago. This synchronicity is likely not a coincidence and may indicate an interconnection between all these objects. The LRD activity seems to be partially responsible for these bursts in the satellites. Assuming that (i) the LRD1 $F410M$ photometry (considering the brightest source of the LRD pair) is contaminated by H$\beta$+[OIII]$\lambda\lambda 4959, 5007$ emission in $\gtrsim25$\% (see Appendix \ref{app:DB}), and (ii) that $\sim10$\% of this composite emission is due to H$\beta$ at this $z$, (iii) the total recombination rate coefficient under Case B, (iv) a hydrogen density of $\sim10^{-4}$ cm$^{-3}$, (v) A(V) $\sim1$ mag, (vi) and a Str\"omgren sphere, we can obtain a rough estimate of the size of the ionizing bubble produced by LRD1 at the time of the observation \citep{Leitherer1995}. We obtained a radius for this bubble on the order of hundreds of pkpc ($\sim500$ pkpc assuming typical SFG line widths and $\sim1$ pMpc if we consider typical AGN line widths), which can thus potentially affect the satellites.

On the other hand, the potential interaction of the LRD pair can also be the trigger for the bursts and later quenching reflected in their SFHs (i.e., environmentally driven bursts, and potential environmental quenching). Given the uncertainties in the SFHs, two possible hypotheses arise for the quenching phase: (a) direct environmental quenching and (b) environmentally triggered internal quenching. These scenarios are depicted in Fig. \ref{fig:scenarios} and potential AGN activity is assumed. The SFHs correspond to those shown in the bottom panels, first row, in Fig. \ref{fig:DB_LRDs_SFH}, which are based on the \texttt{Dense Basis} fits to the stellar component only. As explained in Sect. \ref{sec:sfhs}, these SFHs are compatible with the SFHs obtained from the untouched PSF LRD photometry, which are also displayed for completeness.

In case (a), the interaction can lead to the rapid growth of an AGN (in either of the LRDs) by disrupting the corotational motion of the gas in its vicinity (\citealt{Ricci2017}, \citealt{Davies2022}). Then, the interplay between dust grains' sublimation \citep{Alexander1983} near the AGN and potentially radiation pressure (\citealt{Ricci2017}, \citealt{Arakawa2022}) would leave the immediate surroundings of the supermassive black hole (SMBH) relatively dust-free. Although isolated, the SED shapes of LRD CEERS 746 \citep{Kocevski2023} and CANUCS-LRD-z8.6 \citep{Tripodi2024} suggest the presence of a broad-line AGN possibly caught in a transition phase between a dust-obscured starburst and an unobscured quasar. This stage would have taken place in one or both of our LRDs $\sim100$~Myr ago. If AGN-driven winds are powerful enough, the galaxy might experience a blow-out where large amounts of gas and dust are stripped from the galaxy, potentially suppressing star formation. These outflows could also quench star formation in the companion LRD. The tidal feature shown in Fig. \ref{fig:images} could actually be probing a potential [OII] and [OIII] outflow emission coming from LRD1.

\begin{table*}[htp]
\setlength{\tabcolsep}{2.4pt} 
\renewcommand{\arraystretch}{1.5}
\caption{Hypotheses for this galaxy group.}        
\centering          
\footnotesize
\begin{tabular}{c|c|c|c|c|c}        
\hline\hline                 
  Source & $z$ case &\texttt{Dense Basis} (Stellar fit) & \multicolumn{2}{c|}{Bagpipes (Stars + AGN)}& \texttt{Dense Basis} (Stellar fit)\\ \hline
  & & &Stellar component& AGN& Stellar component\\ 
  \hline 
     & low-$z$ & Dusty and massive BB galaxy & - & - & - \\
 LRD1& mid-$z$ & Dusty and massive BB galaxy & Low-mass unattenuated SFG & Obscured AGN & Low-mass unattenuated SFG\\
  & high-$z$ & Dusty and massive BB galaxy & Dusty and very massive BB galaxy & Dust-free AGN & Dusty and very massive BB galaxy\\
  \hline
 LRD2& ALL & Dusty and massive BB galaxy & Dusty and massive BB galaxy & Dust-free AGN & Dusty and massive BB galaxy\\
 \hline \hline
    & low-$z$ & Massive unattenuated SFG & - & - & -\\
 SAT0& mid-$z$ & Massive unattenuated SFG & - & - & -\\
    & high-$z$ & Massive unattenuated BB galaxy & - & - & -\\
 \hline
 SAT1& ALL & Low-mass unattenuated SFG & - & - & -\\
 \hline
 SAT2& ALL & Low-mass unattenuated SFG & - & - & -\\
\hline
\end{tabular}
\label{tab:summary}

\tablefoot{ALL denotes the three redshift intervals, defined in Sect. \ref{sec:sed}. The low-$z$ interval is the one centered at the photo-$z$ of the galaxy group. BB denotes Balmer break and very massive means $M_\star>10^{11}\,M_\odot$. The values for the stellar masses, attenuations, and SFRs can be found in Table \ref{tab:properties}.}
\end{table*}

In case (b), quenching of the AGN host would be delayed with respect to this SMBH rapid growth. If the gas close to the AGN is not highly disturbed (i.e., the merger is less disruptive), then the AGN's impact is lower \citep{Davies2022}. The host galaxy can benefit from gas inflow and compression due to the interaction, enhanced by the AGN. Quenching in the companion LRD would consequently be caused by mechanisms other than the environment (e.g., stellar feedback, feedback from its own AGN). It is important to acknowledge that, given the environmentally driven nature of the burst, this internal quenching can also be regarded as a consequence of the encounter, with the environment acting as the ultimate triggering phenomenon. The difference in the SFH peaks between the two LRDs in case (b) is, however, $\sim50$ Myr, which may not be large enough to be regarded as a real difference given the resolution of the SED-fitting code (i.e., we could still be facing case (a)).

In either case (a) or (b), this is one of the first galaxy systems in which we can test a potential relationship between possible AGN triggering, merging, and reddening in LRDs, as well as a merging origin. Our findings suggest that environmental effects might be playing an important role already at $z\sim7$. 

It is important to remember that these scenarios are based on the assumption that the Balmer breaks we see are stellar. According to \citet{Ji2025}, a combination of an AGN, a Compton-thick dust-free gas cocoon, and different values of the turbulent broadening can produce smooth Balmer breaks that would be more representative of the breaks seen in the LRD population. This scenario would alleviate the problems associated with the high $M_\star$ and densities of the stellar population that we are seeing. However, the cocoon scenario, while intriguing, is still being explored: the estimated black hole masses are quite high, the formation and evolutionary path of the cocoon are still unclear, not all LRDs show Balmer breaks, and the presence of Balmer absorptions has only been reported in $\sim20$\% of LRDs. This last point could also be due to the lack of high-resolution spectra for these objects, while the absence of these breaks could be related to different evolutionary phases or lower values of the turbulent broadening. Further investigation is required to gain more insight into this possible physical scenario. We note that our LRD pair could naturally fit into this hypothesis, considering that AGN fueling is known to be triggered by host galaxy interactions. We leave the exploration of this possibility to future work.

\section{Conclusions}
\label{sec:conclusions}

We report the discovery of a LRD pair (dubbed LRD1 and LRD2) at $z\sim7$ ($\mathrm{t_{age}} \sim 800$ Myr) in the Abell 370 cluster field using JWST/NIRCam data from CANUCS. This pair forms a compact galaxy group together with three potential satellite sources at the same redshift (SAT0, SAT1, and SAT2). 

According to our SED analysis, both LRDs could be described with a Balmer break plus a high $M_\star$ and A(V). The satellites are dust-free and lower-mass sources. There is a UV excess in the LRDs, especially in LRD1, that cannot be attributed to stars according to current models. We show that this excess could be produced by a potential dust-free AGN, although further confirmation is required. The optical continuum would then be powered by obscured star formation plus AGN emission.

The SFHs of the LRD pair show they recently quenched after a burst of star formation that started $\sim200$ Myr ago, and the satellites have been going through a burst phase that started $\sim100$ Myr ago. The satellites' bursts seem to be triggered by their interaction with the LRDs (i.e., interaction-induced bursts; see \citealt{Asada2023,Asada2024pairs}). On the other hand, the simultaneous burst and quenching of the LRDs could also have an environmental origin, where a potential AGN would shut down the star formation of the whole pair. 

\begin{acknowledgements}
This research was enabled by grant 18JWST-GTO1 from the 
Canadian Space Agency and Discovery Grant and Discovery Accelerator funding 
from the Natural Sciences and Engineering Research Council (NSERC) of Canada to MS.
YA is supported by a Research Fellowship for Young Scientists
from the Japan Society for the Promotion of Science (JSPS). MB
acknowledges support from the ERC Grant FIRSTLIGHT, Slovenian
national research agency ARRS through grants N1-0238 and P1-0188, 
and the program HST-GO-16667, provided through a grant
from the STScI under NASA contract NAS5-26555.
\\
This research used the Canadian Advanced Network For Astronomy 
Research (CANFAR) platform operated in partnership by the Canadian
Astronomy Data Centre and The Digital Research Alliance of Canada
with support from the National Research Council of Canada, the
Canadian Space Agency, CANARIE, and the Canada Foundation
for Innovation.
\\
This research is in part based on observations made with the NASA/ESA \textit{Hubble Space Telescope} obtained from the Space Telescope Science Institute, which is operated by the Association of Universities for Research in Astronomy, Inc., under NASA contract NAS 5–26555. HST observations are associated with program HST-GO-15117. This work is based on observations made with the NASA/ESA/CSA James Webb Space Telescope. The data were obtained from the Mikulski Archive for Space Telescopes at the Space Telescope Science Institute, which is operated by the Association of Universities for Research in Astronomy, Inc., under NASA contract NAS 5-03127 for JWST. JWST observations are associated with program JWST-GTO-1208.
The data described here may be obtained from
\url{https://dx.doi.org/10.17909/xtmj-q319}.
\end{acknowledgements}

% WARNING
%-------------------------------------------------------------------
% Please note that we have included the references to the file aa.dem in
% order to compile it, but we ask you to:
%
% - use BibTeX with the regular commands:
%   \bibliographystyle{aa} % style aa.bst
%   \bibliography{Yourfile} % your references Yourfile.bib
%
% - join the .bib files when you upload your source files
%-------------------------------------------------------------------
%TC:ignore 
\bibliography{reference}
\bibliographystyle{aa}

\begin{appendix}

\section{Lens model}
\label{app:lensing}

\begin{figure}[h]
    \centering
    \includegraphics[width=0.95\linewidth]{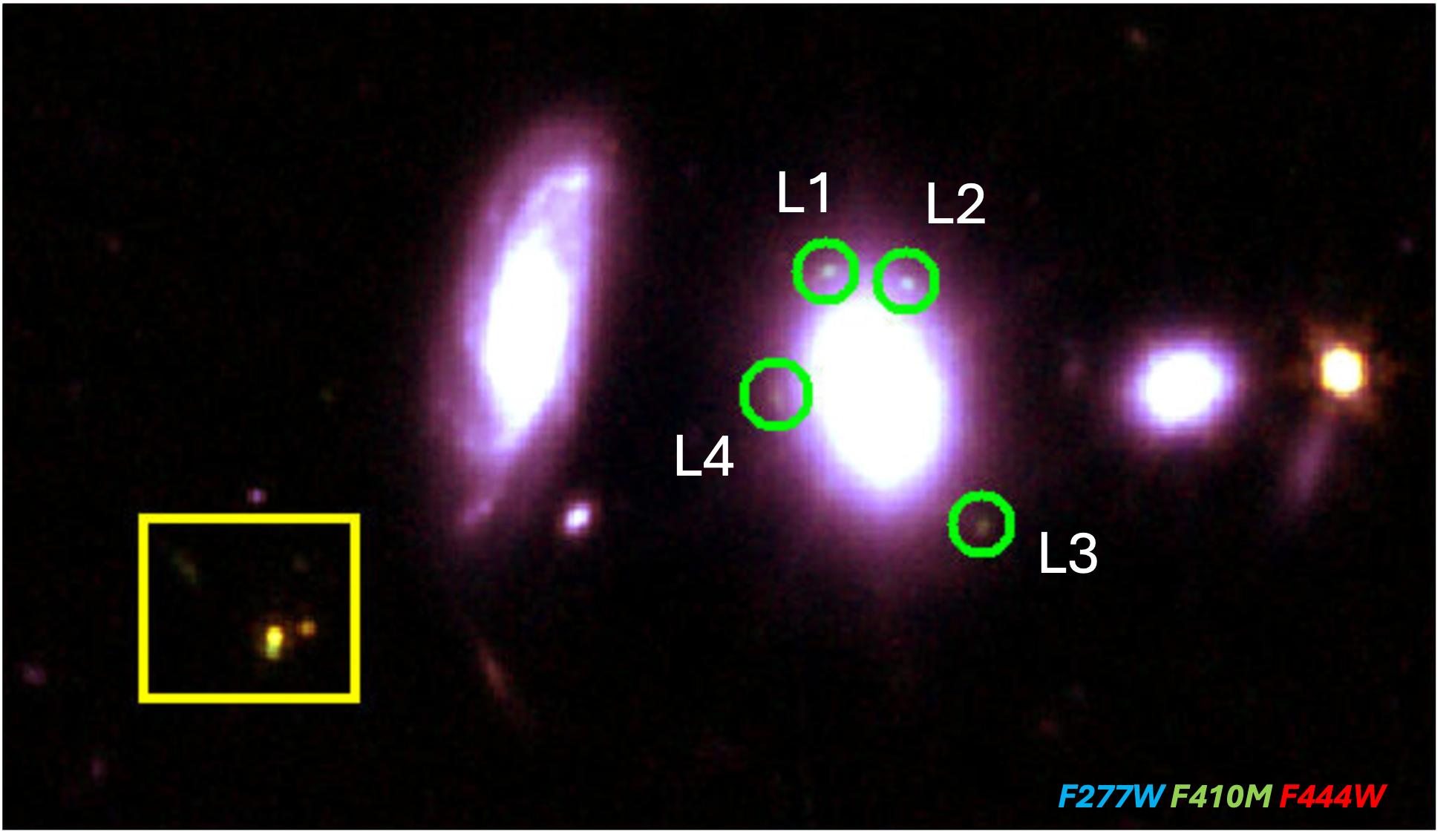}
    \includegraphics[width=0.6\linewidth]{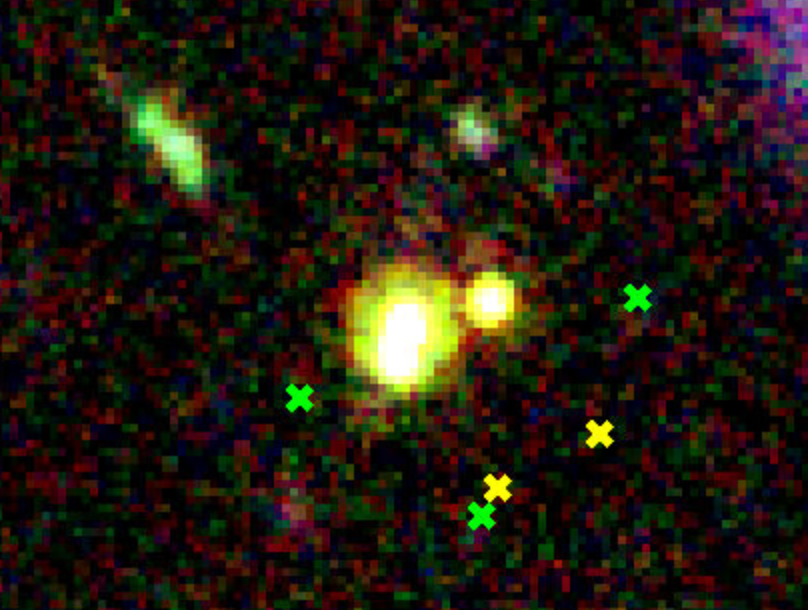}
    \caption{Top: 19$\times$11 arcsec$^2$ cutout of an RGB image based on $F277W$, $F410M$, and $F444W$ centered around coordinates $\alpha=$~2:40:02.338, $\delta=-1$:37:15.297. The yellow square encloses the region that contains our galaxies; its dimensions are $2.8\times2.4$ arcsec$^2$. The four multiple images of the $z\sim7$ background source are highlighted with green circles and labeled with the IDs listed in Table \ref{tab:lens}. The green colors highlight this source's $F410M$ excess; we also noticed  $F410M$ excess in our galaxy sample, suggesting they are at similar redshifts. Bottom: 4$\times$3 arcsec$^2$ cutout of the same RGB image around our sample, with the scale saturated to also show SAT1 and SAT2. The source-plane positions of the sources after the lensing correction are depicted as crosses (yellow for the LRDs and green for the satellites).}
    \label{fig:lensing}
\end{figure}

\citet{Gledhill2024} reported an improved gravitational lensing model for Abell 370 using NIRCam and NIRISS data from CANUCS. 
Additionally, we developed a local lens model for this sample, given that our sources are located over 3\arcmin\, from the cluster center, far from the strong lensing constraints, and close to three bright foreground sources, which were not considered in the \citet{Gledhill2024} strong lensing model. Upon visual inspection, we found that one of the three foreground galaxies (at $\sim 8$\arcsec\, from the LRD pair) features a galaxy-galaxy lensing system and is surrounded by four multiple images of a background source (see Fig. \ref{fig:lensing}). Despite the light contamination from the foreground galaxy, we could identify the background source as an $F090W$ dropout $z\sim 7$ source. We include the positions of the identified multiple images of this source in Table \ref{tab:lens}. As shown in \citet{Desprez2018}, galaxy-galaxy lensing systems on the outskirts of clusters serve as powerful probes of the local lensing properties at large distances where no other strong lensing constraints are available. We therefore used the newly discovered galaxy-galaxy lensing system to constrain our local lens model and accounted for the external shear from the cluster potential. We also note its potential utility for future strong lensing analysis of Abell 370.

This local lens model was constrained with \texttt{Lenstool} (\citealt{Kneib1996}, \citealt{Jullo2007}) and consists of three halos following a dual Pseudo Isothermal Elliptical density profile (\citealt{Limousin2005}, \citealt{Eliasdottir2007}). The elliptical galaxy lensing the background source was modeled with a free central velocity dispersion. The other two galaxies were added as fixed potentials: we modeled the spiral galaxy at 5\arcsec\, away from the LRD pair and the elliptical galaxy at 12\arcsec\, (see Fig. \ref{fig:lensing}), with cluster member scaling relations constrained by \citet{Gledhill2024} in the inner cluster regions, using $F090W$ = 20.14 and 20.89 mag, respectively. The ellipticities and orientations of the three galaxies were fixed, following the light distribution. The cut and core radii of all three galaxies were also obtained from the \citet{Gledhill2024} scaling relations. Following the single lens plane approach, which is conventionally used in cluster lensing analysis (e.g., \citealt{Gledhill2024}), we modeled the sources (with \texttt{EAzY} redshifts between 0.16 and 0.56) at cluster redshift. Apart from the three galaxy halos, we added the external shear with free strength and orientation to account for the contribution of the galaxy cluster. 

\begin{table}[h]
\setlength{\tabcolsep}{0.6pt}
\renewcommand{\arraystretch}{1.5}
    \centering
    \caption{Positions of our galaxies corrected for lensing relative to $\alpha =$ 40.009161, $\delta=-$1.620810 (deg).}
    \begin{tabular}{c|c|c|c}
    Source&CANUCS ID&$\Delta$RA (arcsec)&$\Delta$dec (arcsec)\\
    \hline\hline
       CANUCS-A370-LRD1&2119225&$-7.49$& $-4.08$\\
CANUCS-A370-LRD2&2119226&$-6.99$ &$-3.82$\\
CANUCS-A370-SAT0&2119225&$-7.57$& $-4.23$\\
CANUCS-A370-SAT1&2104864&$-8.48$& $-3.63$\\
CANUCS-A370-SAT2&2104873&$-6.79$& $-3.14$\\
\hline
    \end{tabular}
    \label{tab:lensing_LRD}
    \tablefoot{For each source, we provide the identifier from the original CANUCS catalog.} 
\end{table}

\begin{table}[h]
\setlength{\tabcolsep}{20pt}
\renewcommand{\arraystretch}{1.5}
    \centering
    \caption{IDs and coordinates of the background source's multiple images, lensed by a foreground galaxy near the LRD pair.}
    \begin{tabular}{c|c|c}
    Source&RA&dec\\
    \hline\hline
       L1 &2:40:02.241&$-1$:37:13.374\\
L2 &2:40:02.171& $-1$:37:13.502\\
L3 &2:40:02.104&$-1$:37:16.715\\
L4 &2:40:02.286&$-1$:37:15.017\\
\hline
    \end{tabular}
    
    \label{tab:lens}
\end{table}

\section{Photometry measurements}
\label{app:photometry}
PSF photometry was measured using \texttt{photutils} in PSF-convolved images homogenized to the $F444W$ resolution \citep{Sarrouh2024}. 
We used the empirical PSFs, obtained by median-stacking non-saturated bright stars \citep{Sarrouh2024}. The aperture radius, used to estimate the initial flux of each source, was set to 3 pixels, with fit$\_$shape = (3, 3). Both the fluxes and uncertainties (derived from the residuals of the PSF fit) correspond to the values reported by \texttt{psfphot}. Photometric points were then corrected for Galactic extinction. In Table \ref{tab:phot} we show the photometry measured for each of these galaxies. 

\begin{table*}
\setlength{\tabcolsep}{2.4pt} 
\renewcommand{\arraystretch}{1.5}
\caption{Photometry of this galaxy group.}        
\centering          
\footnotesize
\begin{tabular}{c|c|c|c|c|c|c|c|c|c|c}        
\hline\hline                 
  Source & $F606W$ &$F814W$ &$F090W$ &$F115W$&$F150W$&$F200W$&$F277W$&$F356W$&$F410M$&$F444W$\\ 
  \hline 
  LRD1&8.47$\pm$11.43&9.93$\pm$11.77&10.22$\pm$4.61&45.28$\pm$3.35&53.75$\pm$2.89&54.86$\pm$2.47&81.21$\pm$1.39&346.01$\pm$1.61&517.50$\pm$3.00&471.59$\pm$1.87\\
  LRD2&7.71$\pm$6.87&25.28$\pm$9.05&-&11.26$\pm$3.29&18.37$\pm$1.99&19.85$\pm$2.23&24.21$\pm$1.21&95.92$\pm$1.34&134.81$\pm$2.38&119.13$\pm$2.03\\
  SAT0&-&-&8.70$\pm$4.91&38.82$\pm$11.10&43.17$\pm$11.11&36.74$\pm$7.01&44.75$\pm$4.50&124.62$\pm$5.32&175.30$\pm$8.86&99.73$\pm$7.64\\
  SAT1&$-11.17\pm$13.09&2.07$\pm$14.87&8.07$\pm$3.58&28.27$\pm$3.65&27.98$\pm$3.29&28.19$\pm$2.65&29.97$\pm$2.01&59.18$\pm$2.13&79.86$\pm$3.36&33.18$\pm$2.96\\
  SAT2&$-3.06\pm 6.50$&$-6.60\pm 8.23$&4.15$\pm$1.78&18.28$\pm$1.78&17.23$\pm$1.60&12.07$\pm$1.26&13.64$\pm$0.99&20.75$\pm$1.07&23.84$\pm$1.91&13.53$\pm$1.65\\
\hline
\end{tabular}
\label{tab:phot}

\tablefoot{PSF photometry for LRD1, LRD2, and SAT0. Absent values correspond to those filters in which sources could not be fitted by \texttt{psfphot} due to the high level of noise. SAT1 and SAT2 photometric points were obtained via aperture photometry, extracted from the CANUCS catalogs. All the values were corrected for Galactic extinction and aperture corrections were also applied to aperture photometry. Fluxes and uncertainties are given in nJy units, not corrected for lensing magnification ($\mu=1.26$).}
\end{table*}

\begin{figure*}[htp]
\centering
    \begin{subfigure}{0.75\textwidth}
    \includegraphics[width=\textwidth]{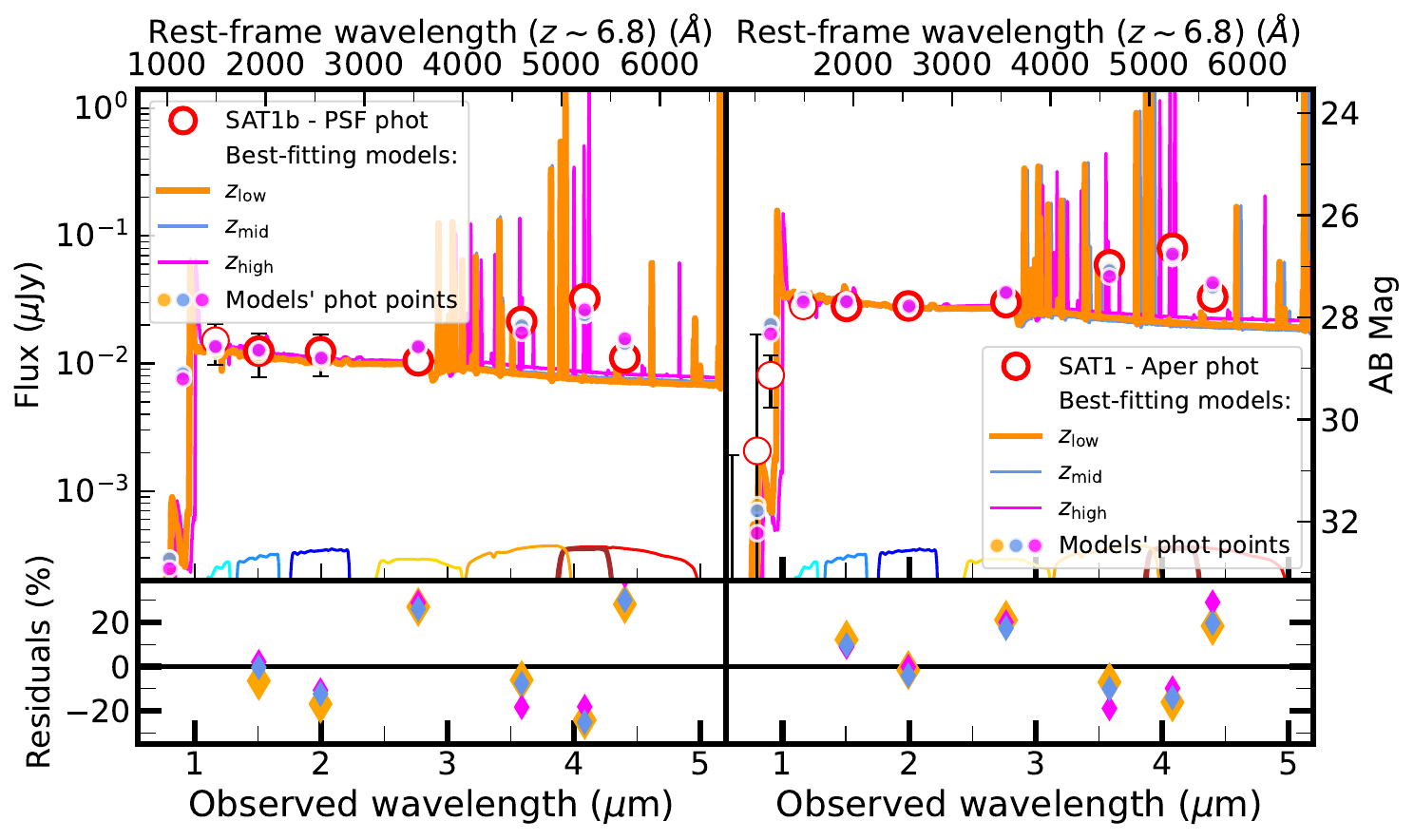}
    \end{subfigure}
    
    \begin{subfigure}{0.75\textwidth}
    \centering
    \includegraphics[width=\textwidth]{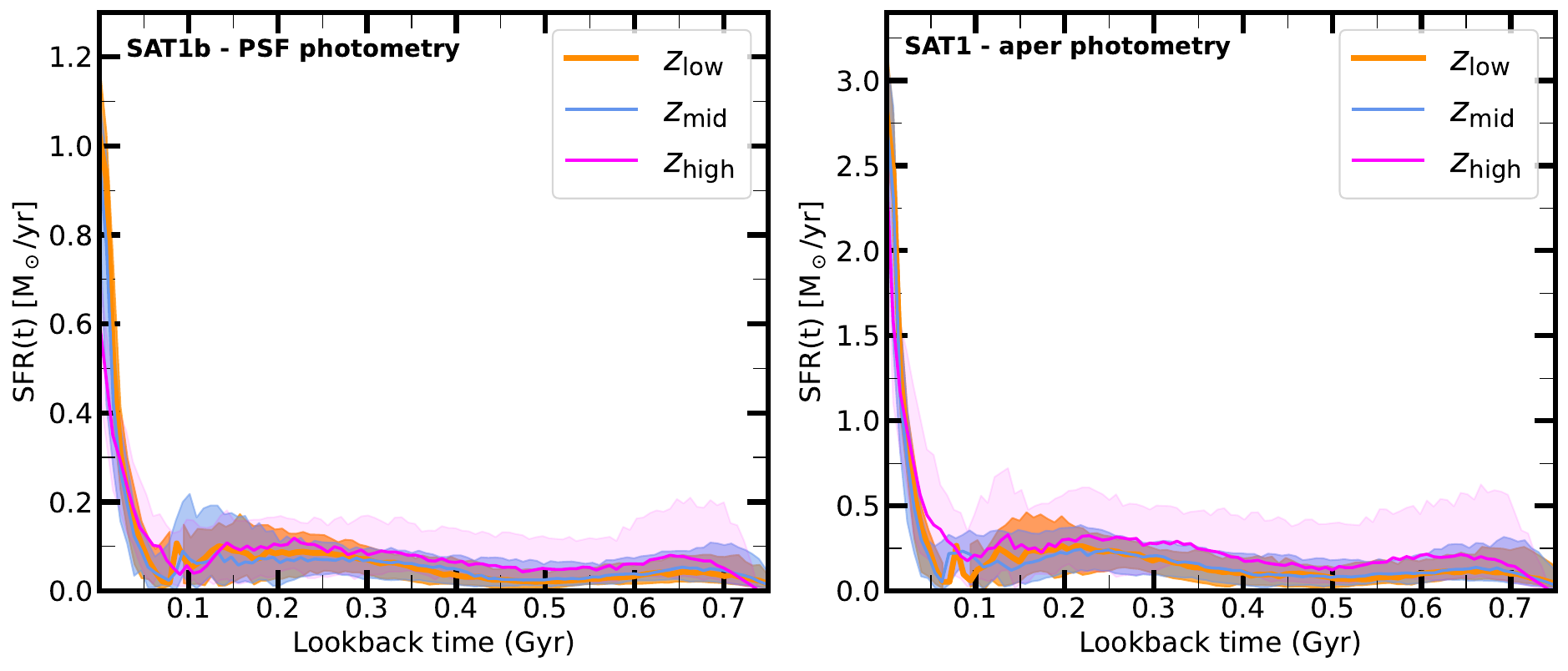}
    
   \raisebox{12.3ex}[0pt][0pt]{\hspace{9.5cm}
   {\includegraphics[width=0.22\textwidth]{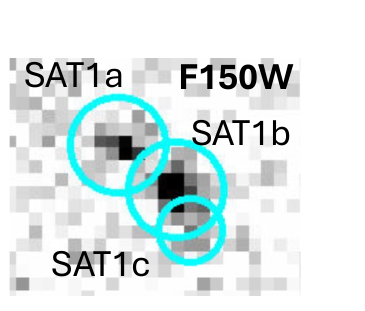}}}
    
    \end{subfigure}
    \caption{\texttt{Dense Basis} stellar fits and SFHs derived for SAT1b and SAT1. We include a $F150W$ cutout of SAT1 that shows its different components in the bottom right panel. See Fig. \ref{fig:DB_LRDs_fits} for a description of the color codes.}
    \label{fig:SAT1b_fits}
\end{figure*}

SAT1, whose analysis in this work is based on aperture photometry, shows signs of a multi-component structure. The main goal of this paper is to try to understand the nature of the LRD pair and determine whether there is evidence of a connection between all the galaxies in this potential group via their SFHs. Understanding the full complexity of SAT1 is beyond the scope of this work. Nevertheless, we offer a possible segmentation for this object, as well as the PSF photometry, \texttt{Dense Basis} SED fitting, and SFHs of its main component, SAT1b, in Fig. \ref{fig:SAT1b_fits}.
The emission of SAT1 seems to be dominated by SAT1b, located at $z\sim6.79$. It is a $\sim10^8\, M_\odot$ source with A(V) $\sim0.1$ mag according to \texttt{Dense Basis}. SAT1a could be a lower-$z$ source ($z\sim5.9$) while SAT1c, which seems to be a clump or satellite of SAT1b, is located at $z\sim6.77$.

\section{SAT3}
\label{app:SAT3}

\begin{figure}[htp]
    \centering
    \begin{subfigure}{0.5\textwidth}
    \includegraphics[width=0.9\linewidth]{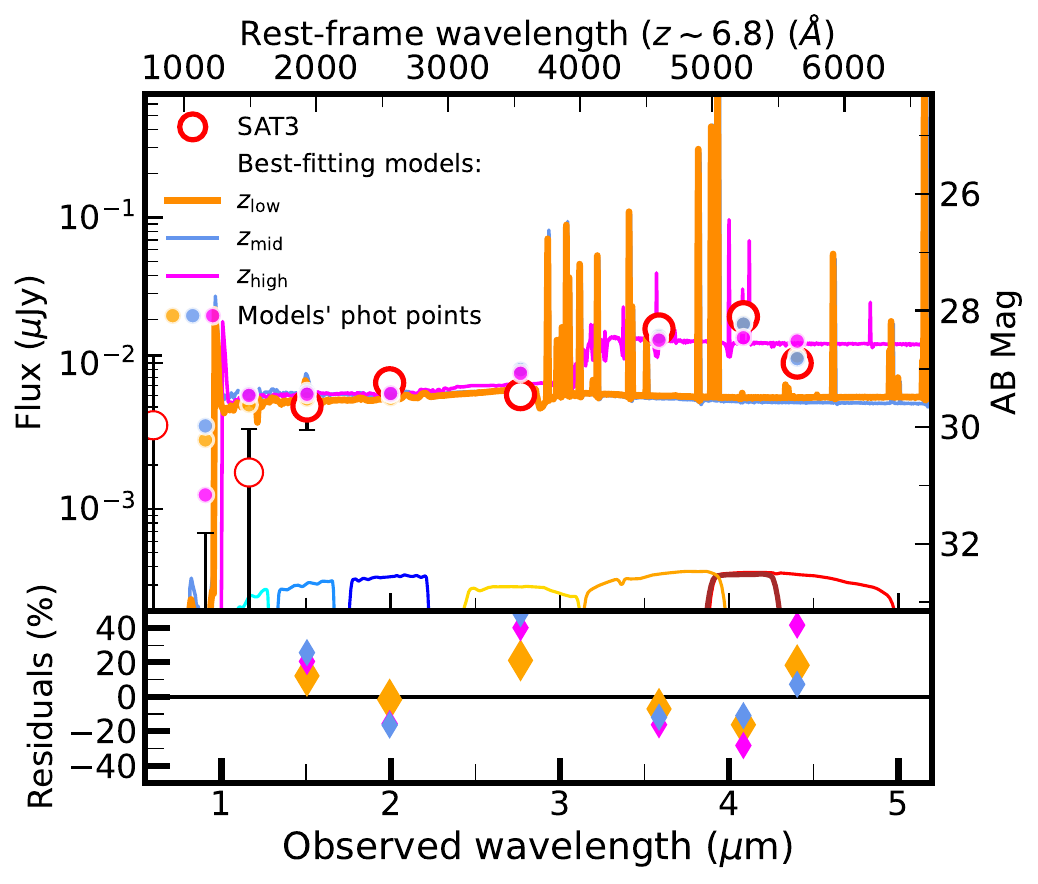}
    \end{subfigure}
    \begin{subfigure}{0.5\textwidth}
    \includegraphics[width=0.8\linewidth]{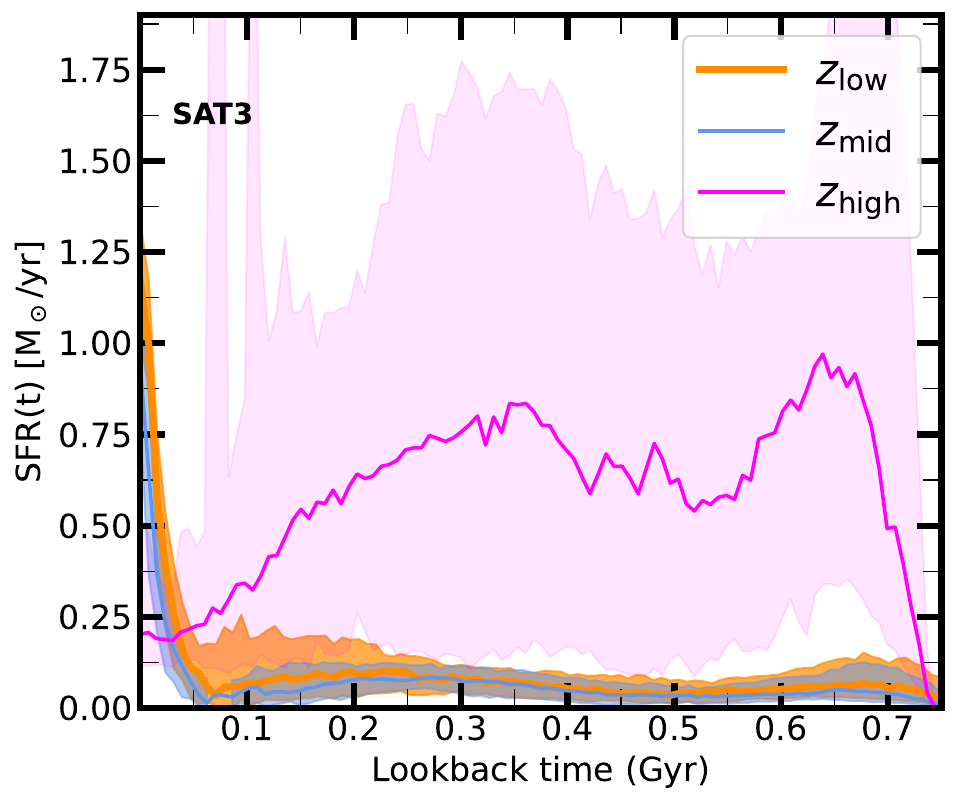}
    \end{subfigure}
    \caption{\texttt{Dense Basis} stellar fits and SFHs for SAT3. See Fig. \ref{fig:DB_LRDs_fits} for a description of the color codes.}
    \label{fig:SAT3_SED}
\end{figure}

The first row in Fig.~\ref{fig:SAT3_SED} shows the 0.3\arcsec\, aperture photometry from the CANUCS catalog of source \#2104938, dubbed SAT3 in Sect. \ref{sec:basic_proprerties}, together with the \texttt{Dense Basis} stellar fits. This galaxy is located at $z\sim6.79$, corresponds to a $10^8\,M_\odot$ source, and is subject to a low dust attenuation (A(V) $\sim 0.2$ mag), showing similar properties to SAT1 and SAT2. The low- and mid-$z$ SED fits point to an SFG, while the high-$z$ solution fits a Balmer break, as also seen in SAT0. As in that case, the colors of this source favor the SFG scenario. The SFHs associated with each of the $z$ intervals are displayed in the second row in Fig. \ref{fig:SAT3_SED}. The low- and mid-$z$ cases depict a recent burst of star formation, similar to those seen in the SFHs of SAT1 and SAT2.

\section{The \texttt{Dense Basis} method}
\label{app:DB}

As specified in Sect. \ref{sec:DB}, for the \texttt{Dense Basis} fits the $M_\star$ was constrained between \mbox{$7<\mathrm{log}\, M_\star/M_\odot<12$} and the SFR between \mbox{$-1<\mathrm{log\, SFR} [M_\odot/\mathrm{yr}]<3$}, setting a flat specific SFR. For parameters such as dust attenuations or metallicities, the default \texttt{Dense Basis} configuration was used, which considers a \cite{Calzetti2000} dust attenuation law. This assumption for the attenuation is reasonable for our galaxies, as the dust curves at $z>6$ are commonly found to be flat and lack a prominent UV bump (\citealt{Markov2023, Markov2024}). In \cite{Tripodi2024} we adopted a flexible analytical attenuation model for an LRD at $z\sim8$. The resulting attenuation curve was Calzetti-like, though slightly shallower in the rest-frame UV. However, changes in the shape of the curve had a considerable impact on the inferred A(V) by $\sim0.6$ dex. This, in turn, affects other fundamental galaxy properties to a lesser extent, such as $M_\star$, SFR, or stellar age by $0.2 - 0.4$ dex. We cannot provide better constraints on the attenuation with our current data and insist on keeping this possible degeneracy in mind.

We selected what we called a low- ($6.75<z<6.85$, centered at the photo-$z$ of the galaxy group), mid- ($6.75<z<6.95$), and high-$z$ ($6.95<z<7.15$) intervals for the redshift priors.

\texttt{Dense Basis} identifies a best-fitting solution for each source looking for the model within the atlas (which is set according to the selected priors) that maximizes the likelihood. Upper and lower bounds for each parameter are then computed based on the 100 best models for each galaxy. 
This approach will penalize those templates that exhibit a slight deviation from the points with the highest S/N (in our case the red points: the $F356W$, $F410M$, and $F444W$ data). Conversely, it will be more flexible in selecting how the lower S/N points are fitted (the blue points: the $F150W$, $F200M$, and $F277W$ data). This is not an issue for the satellites, which are typical SFGs and are easily reproduced by current models. However, LRDs are still not well represented by any known template.

\begin{figure*}[htp]
\centering
    \includegraphics[width=.9\textwidth]{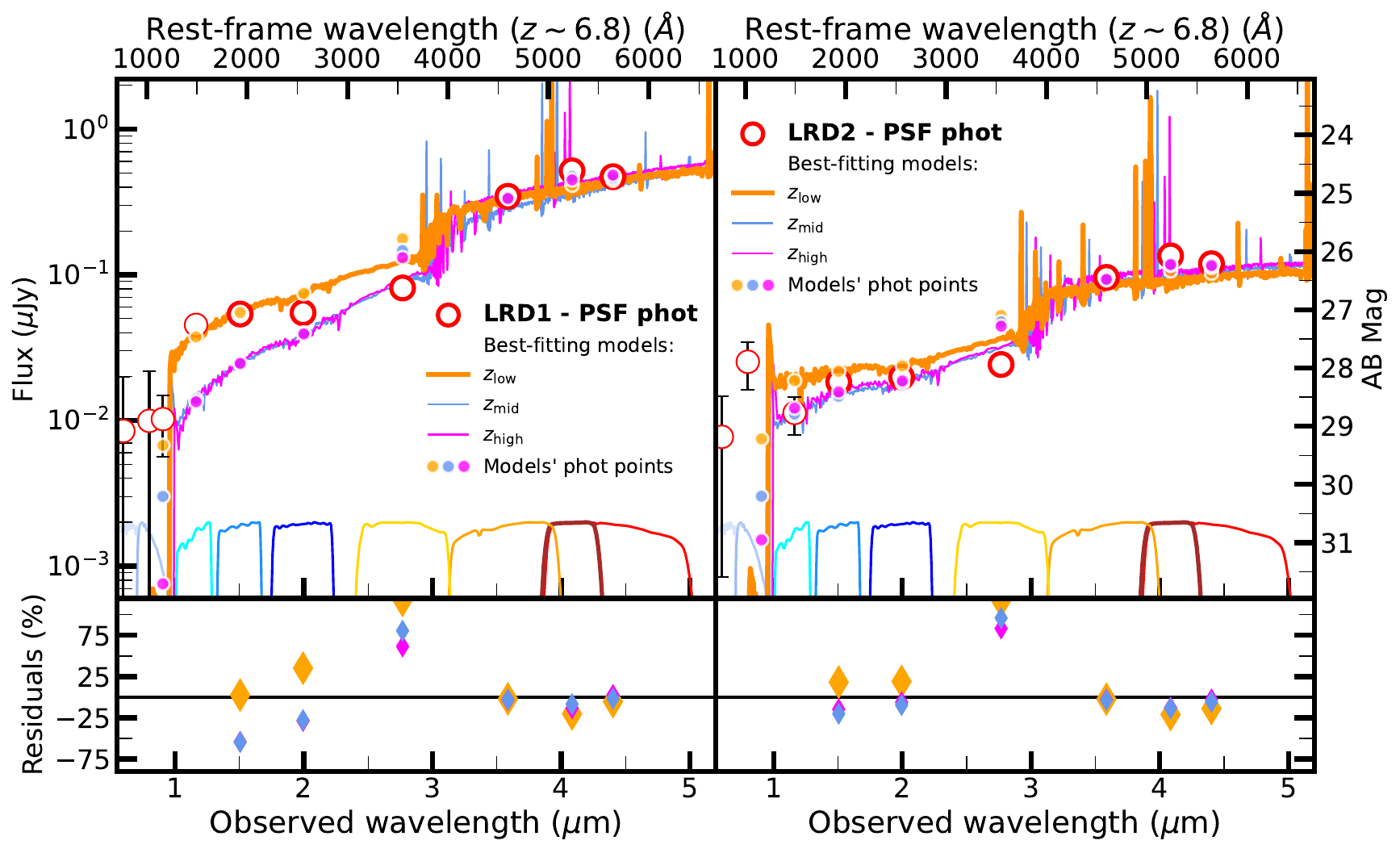}
    
    \caption{Stellar fits obtained directly with \texttt{Dense Basis} based on the LRD1 (left) and LRD2 (right) PSF photometry. See Fig. \ref{fig:DB_LRDs_fits} for a description of the markers and color codes.}
    \label{fig:DB_LRD_fits_0}
\end{figure*}

\begin{figure*}[htp]
\centering
    \includegraphics[width=.9\textwidth]{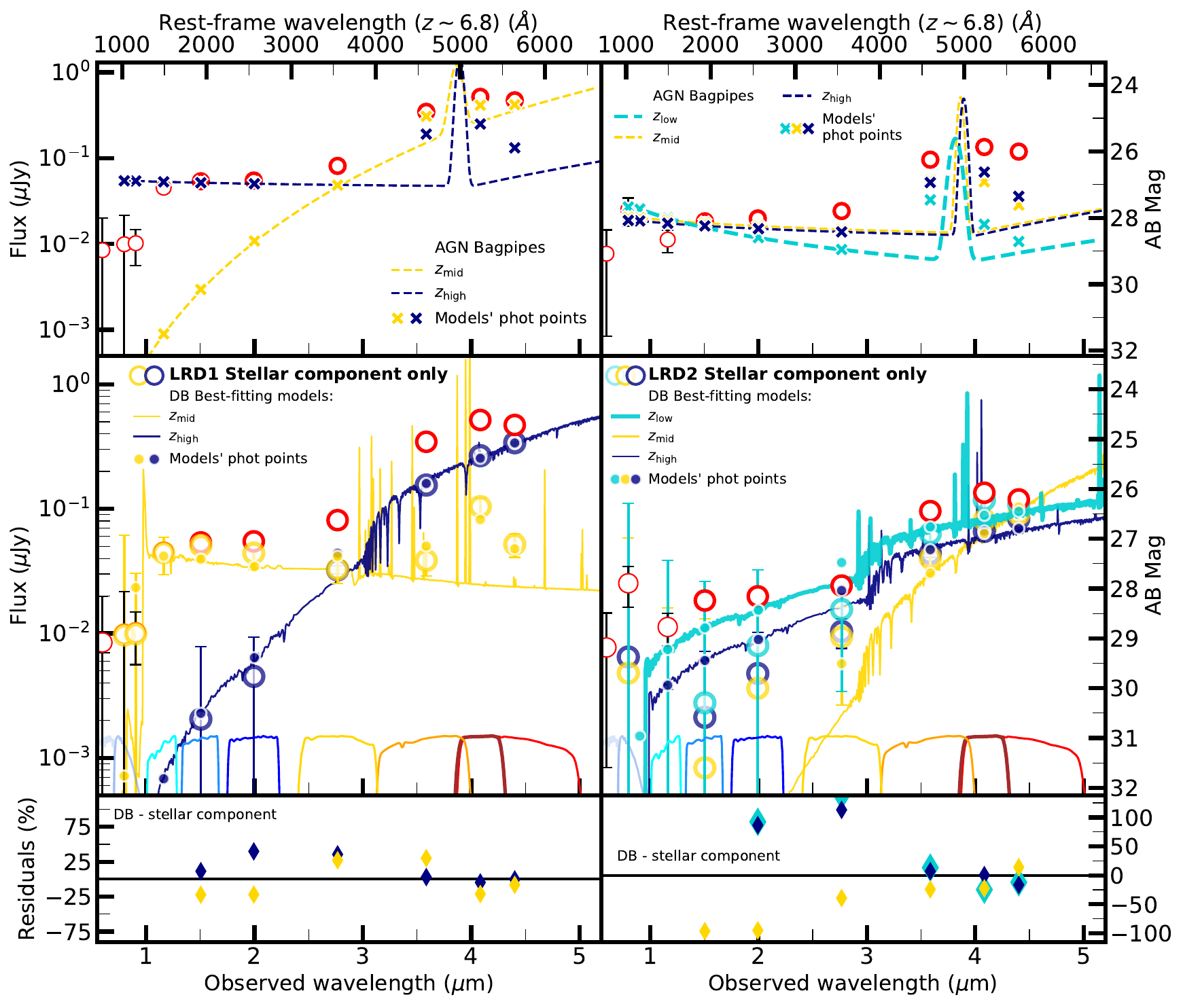}
   
    \caption{Stellar fits obtained directly with \texttt{Dense Basis} based on the stellar component (derived by subtracting the \texttt{Bagpipes} AGN photometry from the PSF photometry) of LRD1 (left) and LRD2 (right) PSF photometry (see Sect. \ref{sec:stellar_component}). See Fig. \ref{fig:DB_BP_LRDs_fits} for a description of the markers and color codes.}
     \label{fig:DB_BP_LRDs_fits_0}
\end{figure*}

Figures \ref{fig:DB_LRD_fits_0} (based on the LRD PSF photometry) and \ref{fig:DB_BP_LRDs_fits_0} (based on the stellar component only) depict the best-fitting models obtained directly with this code for the LRD pair. In some of the cases, \texttt{Dense Basis} yielded best-fitting models that could reproduce the red slopes by selecting a Balmer break, but these models exhibited a UV slope that was either steeper or flatter than the observed one. An illustrative example of such models is provided in the right panel of Fig. \ref{fig:DB_LRD_fits_0} (orange curve) and in the right panel of Fig. \ref{fig:DB_BP_LRDs_fits_0} (yellow curve). 

An alternative option would be to fit the LRDs with two different model components and two different slopes (see \citealt{Killi2024}, \citealt{Setton2024}). However, this approach does not align with the scientific goal of this work, as these fits would not allow us to establish a connection between the photometry and the SFHs.

In an attempt to obtain a model that balances  the fitting of both the red and blue points, we opted to iteratively increase the uncertainties of the red data in some percentage of the flux. This provides more freedom to \texttt{Dense Basis}, which can slightly deviate from the trend defined by the red points while looking for a more suitable solution for the UV. The uncertainties of the red points were increased in $0-2$\% of the flux in 0.05\% steps while keeping the best-fitting model for each iteration. To select the final most probable solution, we compared the models' slopes defined by the continuum around different parts of the SED with those directly measured from the data. We defined a red slope by means of a linear fit to the $F356W$ and $F444W$ data points; a blue slope, using the $F150W$, $F200W$, and $F277W$ filters; and the break, based on the $F277W$ and $F356W$ emission. We kept the 10 best-fitting models that minimized the ratio between the observed and the model's red slopes and then selected the model that minimized the ratios of the blue slopes and the breaks. 

We are aware that this approach will preferentially select Balmer break solutions in which the red points probe the continuum redward of the break instead of models of highly SFGs where these are partially or totally boosted by strong emission lines. Nevertheless, this approach was also used to fit the SEDs of the satellites, resulting in the recovery of star-forming solutions with no breaks. On the other hand, as already mentioned, \texttt{Dense Basis} also favors the presence of a Balmer Break without additional constraints on the code. Furthermore, we also explored selecting a more restrictive prior on the attenuation and SFR to force a highly star-forming solution. We set A(V) $>0.8$ mag and \mbox{$1<$log SFR[$M_\odot$/yr]$<3$} for the priors. However, during the fitting process, \texttt{Dense Basis} rescaled these values and continued selecting Balmer break solutions with lower SFRs.

If we therefore assume that these LRDs have a Balmer break, part of the emission could still be powered by lines contaminating the broadband photometry. This would have a direct impact on our stellar mass estimates, which could be larger than the actual $M_\star$ (see \citealt{Desprez2024}). At the redshifts considered in this work, the $F356W$, $F410M$, and $F444W$ points may be contaminated by the H$\beta$+[OIII]$\lambda\lambda 4959, 5007$ composite emission, whereas the [OII]$\lambda\lambda 3726,3729$ doublet may boost the $F277W$ filter. Only if $z\gtrsim7.0$ we can assume that the $F356W$ filter is free from emission lines, thereby allowing us to probe the continuum. In that case, the three-filter method described in \cite{Vilella-Rojo2015} can be used to measure the contamination of these lines in the $F410M$ and $F444W$ filters, and define the continuum redward of the break. According to this method, the H$\beta$+[OIII]$\lambda\lambda 4959, 5007$ emission would be responsible for $\sim11$\% of the flux in $F444W$ and $\sim25$\% of the flux in $F410M$ in the case of LRD1 ($\sim11$\% and $\sim24$\% in the case of LRD2). Running \texttt{Dense Basis} using the line-corrected fluxes yields similar $M_\star$ compared to the results based on the original photometry, with $\sim0.26$ dex and $\sim0.10$ dex lower $M_\star$ values for LRD1 and LRD2, respectively. However, this correction is quite uncertain, given that we are assuming that the $F356W$ filter is free from e-line emission and that we are also not correcting the $F277W$ filter for possible contamination. Additionally, outshining of older stars by young low-metallicity stars becomes important at $z>7$, which can also lead to bad estimations of the $M_\star$ (\citealt{Narayanan2024}, \citealt{Whitler2023}).
These are all reminders that our results for the $M_\star$ of the LRD pair should only be considered with extreme caution.

\section{The \texttt{Bagpipes} method}
\label{app:bagpipes}

\begin{figure}[htp]
\centering
    \includegraphics[width=0.5\textwidth]
    {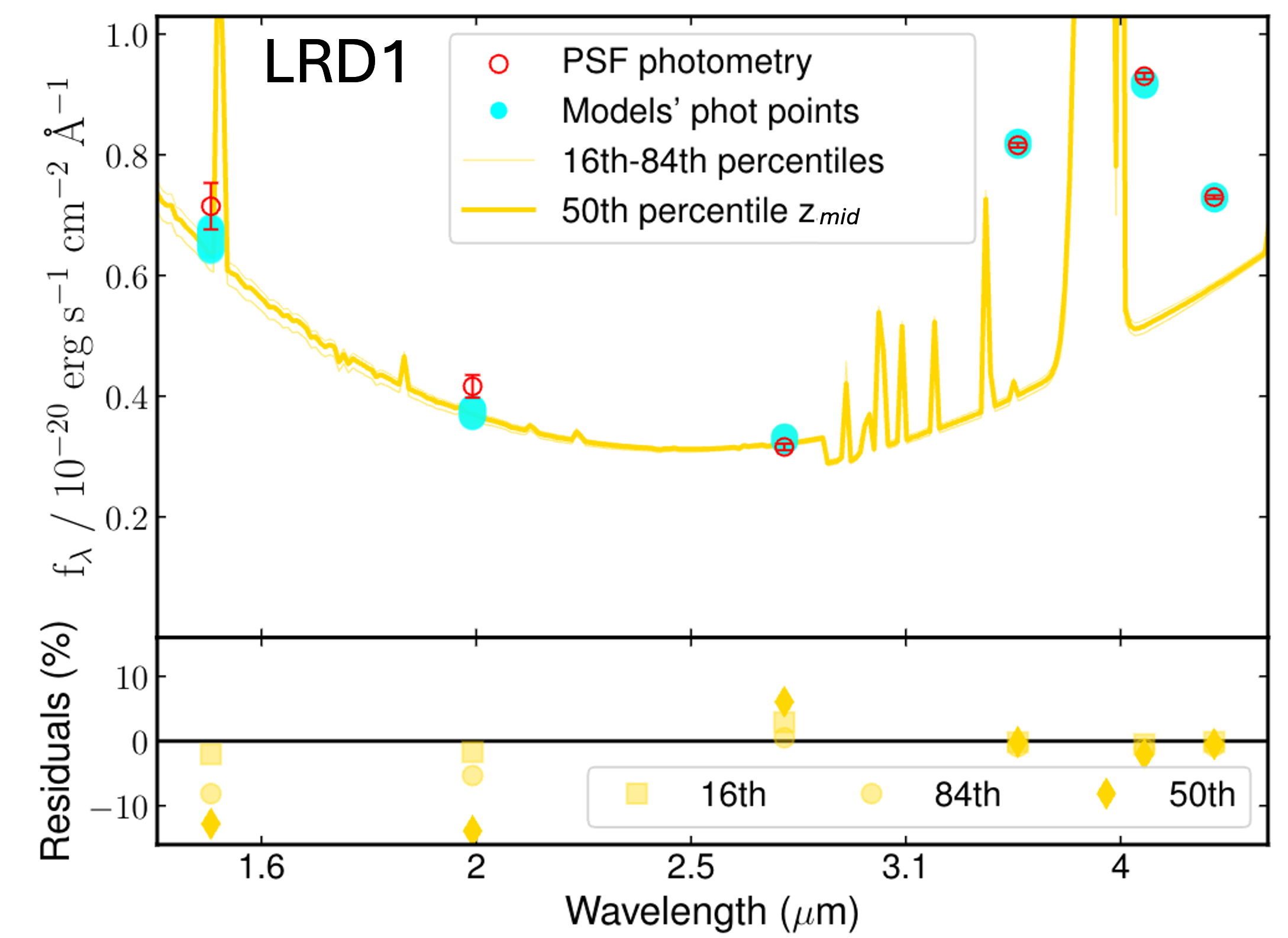}
    \caption{\texttt{Bagpipes} fit (stars+AGN) for LRD1 based on PSF photometry (red open circles), corresponding to the mid-$z$ case. The best-fitting model is shown as a thick yellow line (50th percentile). See Fig. \ref{fig:bagpipes_lrd1_highz} for a full description of the color codes and markers.}
    \label{fig:LRD1_bagpipes_zint}
\end{figure}

\begin{figure}[htp]
\centering
    \includegraphics[width=0.5\textwidth]{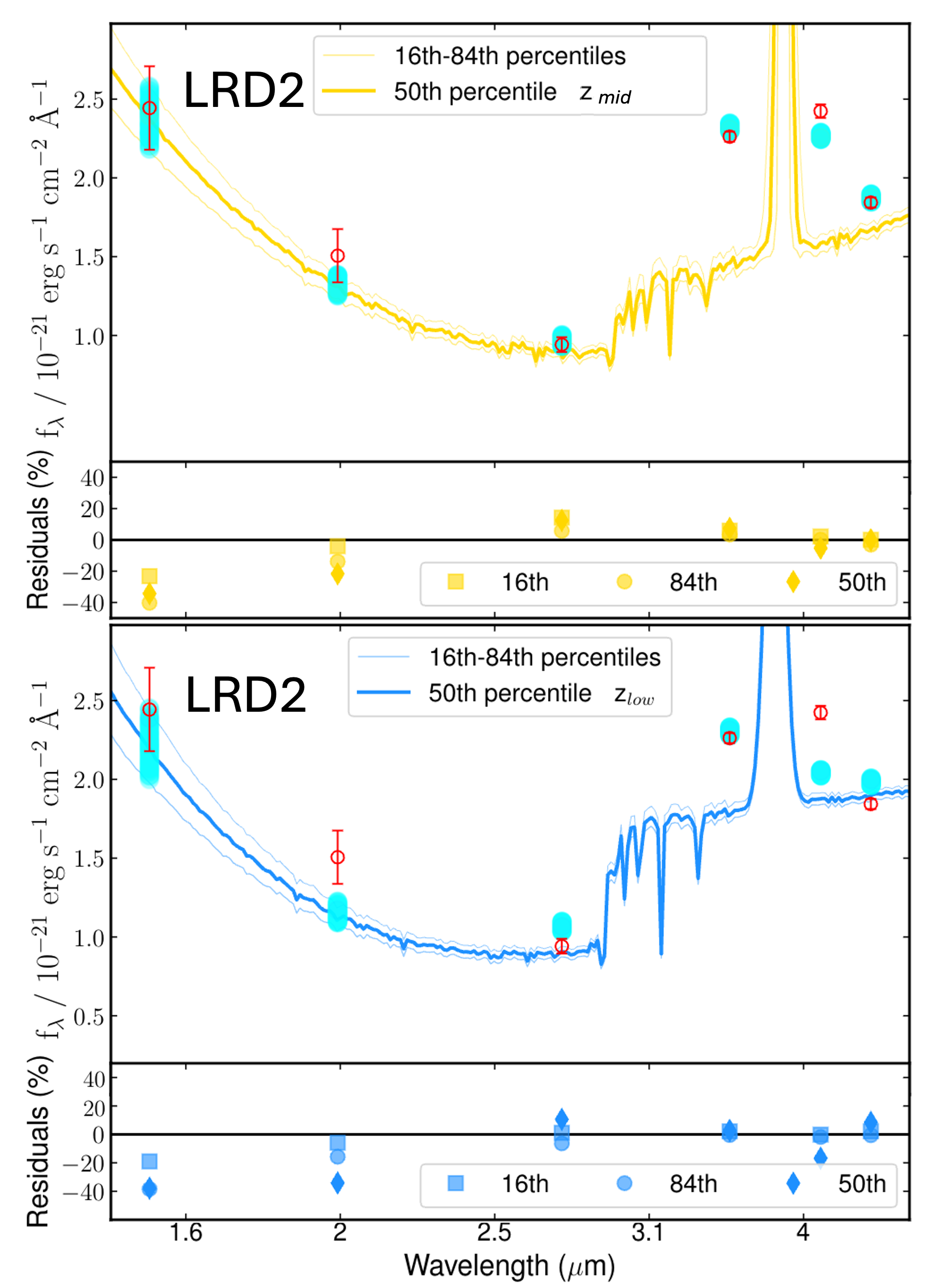}  
    \caption{\texttt{Bagpipes} fit (stars+AGN) for LRD2 based on PSF photometry (red open circles), corresponding to the mid-$z$ (top) and low-$z$ (bottom) cases. Best-fitting models are shown as yellow (top) and blue (bottom) thick lines (50th percentile). See Fig. \ref{fig:bagpipes_lrd1_highz} for a full description of the color codes and markers.}
     \label{fig:LRD2_bagpipes}
\end{figure}

\texttt{Bagpipes} is a \texttt{Python} software that utilizes a Bayesian inference approach to estimate physical parameters of galaxies using spectroscopic and/or photometric data. We fitted the photometric points of the two LRDs using a composite model consisting of a stellar population plus an AGN. Initial tests employing only single AGN components or solely stellar populations did not yield satisfactory results. Specifically, when fitting only the stellar component, \texttt{Bagpipes} encountered similar issues as \texttt{Dense Basis}, identifying a Balmer break galaxy that fails to reproduce the blue part of the spectrum and the flux in the $F277W$ filter. Similarly, a model with only an AGN component failed to account for the blue part of the spectrum due to the limitations of the rising power-law continuum. All fits were performed using the Multinest sampling algorithm. 

The composite model comprises several parameters: 

\begin{itemize}
    \item The AGN continuum emission was modeled as a broken power law characterized by three parameters: two spectral slopes and the continuum flux at the break point (5100 $\AA$). 
    \item The AGN emission line included as a single H$_\beta$ emission line, modeled with two free parameters: line width and line intensity. 
    \item The SFH was modeled using a double power law, enabling the representation of both quenched galaxies and SFGs while minimizing the number of free parameters. The free parameters include the two slopes, the decay timescale ($\tau$), the stellar mass formed, metallicity, and the time when the stars started to form (age). 
    \item Nebular emission was incorporated via the ionization parameter U.
    \item We modeled the dust extinction with the visual extinction parameter A(V), following the Calzetti attenuation law. A dust emission component was not included, as significant emission from hot dust is only expected for $\lambda$ < 1 $\mu$m. 
\end{itemize}

Note that the emission due to the [OIII]$\lambda\lambda$4959, 5007 lines is included through the ionization parameter U \citep{Boyler2017}. In fact, these lines are present in the mid-$z$ solution shown in Fig. \ref{fig:LRD1_bagpipes_zint}. However, this emission is indistinguishable from the H$\beta$ line in our photometric data, i.e., Bagpipes cannot accurately assign the corresponding separate fluxes to [OIII] and H$\beta$. We would need a better sampling of the SED to carry out a more detailed and complex modeling of these galaxies.

The redshift parameter was treated in two ways. First, we applied the same prior intervals used in the \texttt{Dense Basis} fits (high, mid, and low-$z$ priors, as shown in Table \ref{tab:parameters}). Second, we employed a broader uniform prior ranging $6.7<z<7.1$, encompassing all the other intervals. For LRD1, the broader prior produced a Gaussian posterior distribution consistent with the high-$z$ prior, where the 50th percentile aligns with the posterior peak (see Fig. \ref{fig:bagpipes_lrd1_highz}). In contrast, the mid-$z$ prior (Fig. \ref{fig:LRD1_bagpipes_zint}) resulted in a non-Gaussian posterior peaking beyond the upper limit, causing a discrepancy between the 50th percentile and the peak. For the mid-$z$ prior, the AGN parameters' posterior distributions also lose their Gaussian shape. For the low-$z$ prior, \texttt{Bagpipes} failed to converge to a solution with the same parameter set. 

For LRD2, the redshift behavior was similar. The high-$z$ prior yielded a Gaussian posterior (Fig. \ref{fig:bagpipes_lrd1_highz}), while the mid and low-$z$ priors still peaked around $z\sim7$ (Fig. \ref{fig:LRD2_bagpipes}). However, \texttt{Bagpipes} successfully converged for the low-$z$ interval in this case. 

All model parameters, along with their prior distributions and posterior 50th percentile values, are summarized in Table \ref{tab:parameters} for both LRDs. The prior distributions and limits follow those used by \cite{Tripodi2024}.

\begin{table*}[htp]
\centering
\caption{Model parameters, priors, and 50th percentile posterior values for LRD1 and LRD2.}
\begin{tabular}{l|c c c|c c c c}
\hline
\multirow{2}{*}{Parameter} & \multicolumn{3}{c|}{LRD1} & \multicolumn{4}{c}{LRD2} \\ \cline{2-8}
 & Prior & $z_{high} fit$ value & $z_{mid} fit$ value & Prior & $z_{high} fit$ value & $z_{mid} fit$ value & $z_{low} fit$ value \\ 
\hline\hline
Redshift: $z_{high} fit$ & 6.95 - 7.15 & $7.01^{\,0.1}_{\,-0.1}$ & $-$ & 6.95 - 7.15  & $7.0^{\,0.01}_{-0.01}$ & $-$ & $-$\\[0.9ex]
\phantom{Redshift:} $z_{med} fit$ & 6.75 - 6.95 & $-$ & $6.95^{\,0}_{0}$ & 6.75 - 6.95 & $-$ & $6.95^{\,0.0}_{-0.}$ & $-$\\
\phantom{Redshift:,}$z_{low} fit$ & $-$ & $-$ & $-$ & 6.75 - 6.85 & $-$ & $-$ & $6.85^{\,0.}_{\,-0.}$ \\[0.9ex]
\hline
AGN: $\alpha_\lambda$ & ($-4$, 2) & $-2.09^{\,+0.16}_{\,-0.18}$& $2.56^{\,0.04}_{-0.02}$ & ($-4$, 2) & $-2.29^{\,0.25}_{-0.24}$ & $-2.28^{\,0.22}_{-0.22}$ & $-2.79^{\,0.24}_{\,-0.14}$ \\[0.9ex]
AGN: $\beta_\lambda$ & ($-2$, 3) & $0.55^{\,+0.46}_{\,-0.46}$ & $1.88^{\,+0.17}_{-0.16}$ & ($-2$, 2) & $0.53^{\,0.47}_{-0.44}$ & $0.47^{\,0.47}_{-0.46}$ & $0.45^{\,0.44}_{\,-0.43}$ \\[0.9ex]
AGN: $\sigma$ & (30, 4200) & $2400^{\,+2}_{\,-2}$ & $3802^{\,1}_{-1}$ & (30, 4200) & $1995^{\,2}_{-2}$ & $2137^{\,2}_{-3}$ & $3311^{\,2}_{-2}$ \\[0.9ex]
\hline
SFH: $\alpha$ & (0.01, 1000.) & $79^{\,+6}_{\,-6}$ & $1.02^{\,30}_{\,-17}$ & (0.01, 1000.) & $78^{\,5}_{-6}$ & $69^{\,6}_{-6}$ & $0.22^{\,2}_{-2}$ \\[0.9ex]
SFH: $\beta$ & (0.01, 1000.) & $0.8^{\,+55}_{\,-18}$ & $871^{\,3}_{-1}$ & (0.01, 1000.) & $5^{\,25}_{-10}$ & $2^{\,27}_{-8}$ & $1.4^{\,1.04}_{-0.82}$ \\[0.9ex]
SFH: $\tau$ & (0.01, 15.) & $0.2^{\,+0.16}_{\,-0.12}$ &  $0.76^{\,0}_{0}$ & (0.01, 15.) & $0.22^{\,0.19}_{-0.14}$ & $0.08^{\,0.12}_{-0.05}$ & $0.45^{\,0.10}_{-0.12}$ \\[0.9ex]
SFH: log($M_{\star}/M_\odot$) & (7, 12) & $11.11^{\,+0.07}_{\,-0.08}$ & $7.56^{\,0.04}_{0.03}$ & (7, 12) & $10.5^{\,0.09}_{-0.1}$ & $10.6^{\,0.05}_{-0.07}$ & $10.01^{\,0.9}_{-0.09}$ \\[0.9ex]
SFH: $Z/Z_\odot$ & ($-2$, 0.5) & $0.35^{\,+0.1}_{-0.14}$ & $0.10^{\,0}_{0}$ & ($-2$, 0.2) & $-0.78^{\,0.84}_{-0.81}$ & $0.13^{\,0.05}_{-0.08}$ & $0.19^{\,0.0}_{-0.09}$\\[0.9ex]
\hline
Nebular: log(U) & ($-3$, 0) & $-1.05^{\,+0.65}_{\,-0.67}$ & $-0.51^{\,0.25}_{0.27}$ & ($-3$, 0) & $-1.55^{\,0.97}_{-0.89}$ & $-1.52^{\,0.98}_{-0.97}$ & $-1.61^{\,0.21}_{-0.25}$ \\[0.9ex]
\hline
Dust: A(V) [mag]& (0, 2.5) & $2.29^{\,+0.15}_{\,- 0.23}$ & $0.06^{\,0.05}_{0.04}$ & (0, 2.5) & $2.34^{\,0.11}_{-0.19}$ & $2.4^{\,0.07}_{-0.12}$ & $2.46^{\,0.03}_{-0.05}$ \\[0.9ex]
\hline
\end{tabular}
\label{tab:parameters}

\tablefoot{The uncertainties correspond to the values provided by \texttt{Bagpipes}.}
\end{table*}

\section{Physical properties}
\label{app:table}
We report the physical parameters derived with \texttt{Dense Basis}
and \texttt{Bagpipes} for our galaxy system in Table \ref{tab:properties}.

\begin{sidewaystable*}[htp]
\setlength{\tabcolsep}{2.6pt} 
\renewcommand{\arraystretch}{2}
\caption{Physical properties of the galaxy group.}       
\label{tab:properties}    
\centering          
\scriptsize
\begin{tabular}{c| c| c|c| c| c c c c| c c c c|c c c c}        
\hline\hline 
&&&&&&&\texttt{Dense Basis} (stellar fit)&&&&\texttt{Bagpipes} (stars + AGN)&&&&\texttt{Dense Basis} (stellar fit)&\\
\hline
  Source &$\alpha$ &$\delta$ &$z_{\mathrm{EAzY}}$&Redshift prior &&&&&& &&&&&\textbf{Stellar component only}\\ &&&&&$z$&log $M_\star/M_\odot$ & log SFR [$M_\odot$/yr] & A(V) mag & $z$& log $M_\star/M_\odot$ & log SFR [$M_\odot$/yr] & A(V) mag&$z$& log $M_\star/M_\odot$ & log SFR [$M_\odot$/yr] & A(V) mag \\     
\hline

&&&&$6.75<z<6.85$&6.84 &10.41$_{10.41}^{10.41}$ &0.60$_{0.60}^{0.61}$ &1.06$_{1.06}^{1.06}$ &- &- &- &-&- &- &- &- \\
\textbf{LRD1}& 2:40:02.7296 &$-1$:37:18.202& 6.87&$6.75<z<6.95$&6.95&10.45$_{10.45}^{10.45}$&1.03$_{1.03}^{1.04}$ &1.21$_{1.21}^{1.21}$ & 6.95 & $7.56^{7.60}_{7.53}$ & - & $0.06^{0.11}_{0.10}$  &6.95 &8.22$_{7.99}^{8.55}$&0.38$_{0.30}^{0.49}$ & 0.07$_{0.02}^{0.17}$ \\
&&&&$6.95<z<7.15$& 7.14& 10.54$_{10.54}^{10.54}$& $-0.88_{-0.88}^{-0.88}$&1.01$_{1.01}^{1.01}$ & 7.01& $11.11^{11.18}_{11.03}$ & - & $2.29^{2.44}_{2.06}$ &7.14 &11.15$_{11.14}^{11.16}$ & $0.13_{0.07}^{0.90}$&2.77$_{2.77}^{2.85}$ \\
\hline
&&&&$6.75<z<6.85$&6.84 &9.84$_{9.79}^{9.88}$ &0.62$_{0.11}^{0.90}$ &1.26$_{1.03}^{1.44}$ & 6.85& $10.01^{10.10}_{9.92}$ & - & $2.46^{2.49}_{2.41}$ &6.82 &10.29$_{10.23}^{10.36}$ &$-0.03_{-0.96}^{1.06}$ &$2.17_{1.91}^{2.50}$ \\
\textbf{LRD2} &2:40:02.7014 &$-1$:37:18.0550& 6.79&$6.75<z<6.95$&6.95&9.69$_{9.66}^{9.72}$&0.20$_{-0.23}^{0.50}$ &$0.72_{0.57}^{0.90}$ & 6.95& $10.60^{10.65}_{10.53}$ & - & $2.40^{2.47}_{2.28}$ &6.88 &10.44$_{10.36}^{10.52}$&$-0.14_{-1.02}^{1.37}$ & 2.73$_{2.36}^{2.99}$ \\
&&&&$6.95<z<7.15$& 7.13& 9.73$_{9.66}^{9.78}$& $-0.84_{-1.65}^{-0.19}$&$0.62_{0.45}^{0.79}$ & 7.0 & $10.50^{10.59}_{10.40}$ & - & $2.34^{2.45}_{2.15}$ & 7.07 &10.10$_{10.00}^{10.16}$ & $-0.58_{-1.33}^{0.35}$&1.88$_{1.58}^{2.12}$ \\
\hline
&&&&$6.75<z<6.85$&6.81 &9.21$_{8.69}^{9.38}$ &0.48$_{0.00}^{0.81}$ &0.30$_{0.08}^{0.48}$ &- &- &- &-&- &- &- &- \\
\textbf{SAT0}& 2:40:02.7320& $-1$:37:18.355& 6.80&$6.75<z<6.95$&6.88&$9.31_{8.78}^{9.45}$&$0.44_{-0.25}^{0.85}$ &$0.34_{0.08}^{0.56}$ &- &-&-&-&- &-&-&- \\
&&&&$6.95<z<7.15$& 7.07& $9.35_{9.14}^{9.44}$& $0.26_{-0.20}^{0.61}$&$0.14_{0.04}^{0.34}$ & -&- & -&- &- &-&-&- \\
\hline
&&&&$6.75<z<6.85$&6.81 &$8.12_{7.92}^{8.42}$ &$0.32_{0.24}^{0.40}$ &0.06$_{0.02}^{0.15}$ &- &- &- &- &- &-&-&-\\
\textbf{SAT1} & 2:40:02.8072&$-1$:37:17.2690& 6.78&$6.75<z<6.95$&6.87&$8.17_{7.93}^{8.50}$&$0.31_{0.23}^{0.41}$ &$0.07_{0.02}^{0.17}$ &-&-&- &- &- &-&-&-\\
&&&&$6.95<z<7.15$& 7.01& $8.52_{8.13}^{8.86}$& $0.32_{0.14}^{0.51}$&0.12$_{0.03}^{0.32}$ &- &- & -&- &- &- &- &-\\
\hline
&&&&$6.75<z<6.85$&6.80 &$7.96_{7.67}^{8.28}$ &$-0.03_{-0.17}^{0.11}$&$0.09_{0.02}^{0.23}$ &- &- &- &- &- &-&-&-\\
\textbf{SAT2}& 2:40:02.7060&$-1$:37:17.2210&6.79& $6.75<z<6.95$&6.87&$7.97_{7.68}^{8.28}$&$-0.01_{-0.17}^{0.12}$ &$0.10_{0.02}^{0.23}$ &- &-&- & - &- &-&-&-\\
&&&&$6.95<z<7.15$& 7.04& $8.07_{7.77}^{8.34}$& $0.002_{-0.168}^{0.152}$&$0.12_{0.03}^{0.26}$ &- & -&-&- &- &-&-&-\\
\hline                                   
\end{tabular}

\tablefoot{For each source, we provide the physical parameters derived for each of the redshift intervals considered in this work, corrected for lensing magnification ($\mu=1.26$). We distinguish between the results obtained from the PSF/aperture photometry with \texttt{Dense Basis} (stellar fit; Sect. \ref{sec:DB}) and \texttt{Bagpipes} (stellar fit + AGN; Sect. \ref{sec:bagpipes}), and those obtained for the stellar component with \texttt{Dense Basis} (Sect. \ref{sec:stellar_component}), obtained by subtracting the \texttt{Bagpipes} AGN photometric points from the LRD PSF photometry. Upper and lower bounds of each value (which corresponds to the median) indicate the 16th and 84th percentiles. For some values of the \texttt{Dense basis} (stellar fit) of LRD1, both percentiles are identical to the median value, highlighting that only a few models can fit this galaxy using only stars. Coordinates are not corrected for lensing magnification (check Table \ref{tab:lensing_LRD} for the corrected positions).}      
\end{sidewaystable*}

\end{appendix}
%TC:endignore 

%\begin{thebibliography}{}
%
%  \bibitem[Baker(1966)]{baker} Baker, N. 1966,
%      in Stellar Evolution,
%      ed.\ R. F. Stein,\& A. G. W. Cameron
%      (Plenum, New York) 333
%
%   \bibitem[Balluch(1988)]{balluch} Balluch, M. 1988,
%      A\&A, 200, 58
%\end{thebibliography}

\end{document}